\global\def\draftcontrol{0}

   \def\versionno{ Rigid Stabilization }
\catcode`\@=11

\expandafter\ifx\csname draftcontrol\endcsname\relax\global\def\draftcontrol{0}
\fi

{\count255=\time\divide\count255 by 60
\xdef\hourmin{\number\count255}
\multiply\count255 by-60\advance\count255 by\time
\xdef\hourmin{\hourmin:\ifnum\count255<10 0\fi\the\count255}}
\def\draftdate{\number\month/\number\day/\number\year\ \ \ \hourmin }

\newcommand\makepapertitle{\par

  \begingroup
    \renewcommand\thefootnote{\@fnsymbol\c@footnote}%
    \def\@makefnmark{\rlap{\@textsuperscript{\normalfont\@thefnmark}}}%
    \long\def\@makefntext##1{\parindent 1em\noindent
            \hb@xt@1.8em{%
                \hss\@textsuperscript{\normalfont\@thefnmark}}##1}%
     \newpage
     \global\@topnum\z@   % Prevents figures from going at top of page.
     \@makepapertitle
     \thispagestyle{empty}\@thanks
  \endgroup
  \setcounter{footnote}{0}%
  \global\let\thanks\relax
  \global\let\makepapertitle\relax
  \global\let\@makepapertitle\relax
  \global\let\@thanks\@empty
  \global\let\@author\@empty
  \global\let\@date\@empty
  \global\let\@title\@empty
  \global\let\title\relax
  \global\let\author\relax
  \global\let\date\relax
  \global\let\and\relax
  \def\version{\let\version\@version\@gobble}
}
\def\@makepapertitle{%
  \newpage
   \ifnum\draftcontrol=1 {}
   \version\versionno
   \vskip 5em%
   \else
   \hfill\hbox to 3cm {\parbox{4cm}{\@pubnum}\hss}%
   \vskip 5em%
   \fi
   \begin{center}%
   \let \footnote \thanks
      {\hskip -0\textwidth \hbox to 1\textwidth%
        {\centerline{\Large\bf{\noindent\@title}}}}%
     \vskip 2em%
     {\normalsize%\large
       \lineskip .5em%
       \begin{tabular}[t]{c}%
         \@author
       \end{tabular}\par}%
     \vskip 1em%
     {\@bstract}%
     \end{center}%
     \vfill
     \@date%
     \vskip 1.5em%
   \par
}

\gdef\@pubnum{}
\def\pubnum#1{%
  \gdef\@pubnum{#1}}

\gdef\@bstract{}
\def\Abstract#1{%
  \gdef\@bstract{%
   \parbox{\textwidth-0pc}{%
   \centerline{\bf Abstract}\penalty1000
   \noindent%\abstractfont \baselineskip=12pt
   \renewcommand\baselinestretch{1.0}
   {#1}}}
}

\gdef\@email{}
\def\email#1{%
   \gdef\@email{%
   Email: {\tt #1}}
}

\def\ps@paper{\let\@mkboth\@gobbletwo%
     \ifnum\draftcontrol=1
        \def\@oddfoot{\hbox to \textwidth{\tiny \versionno \hfil\tiny\draftdate}%
        \hskip -\textwidth \hbox to \textwidth{\hfil\rm\thepage\hfil}}%
     \else\def\@oddfoot{\hbox to \textwidth{\hfil\rm\thepage\hfil}}
     \fi
     \let\@evenfoot\@oddfoot
}
\def\body{\clearpage
          \pagestyle{paper}
        }
\newenvironment{acknowledgments}{%
\vskip 3.25ex
\noindent {\bf Acknowledgments}
}

\def\@version#1{\ifnum\draftcontrol=1
\typeout{}\typeout{#1}\typeout{}
\vskip3mm\centerline{\hbox{\fbox{\normalsize{\tt DRAFT -- #1 -- }
                   {\draftdate}}}}\vskip3mm
\fi}
\let\version\@version
\long\def\eqlabel#1{\ifnum\draftcontrol=1
                    \tag@false  % there are some problems with multline without this
                    \tag*{(\theequation) \hbox to -0.2cm{\hspace{0cm}\small{#1}\hss}}
                    \refstepcounter{equation}
                    \edef\@currentlabel{\theequation}
                    \ltx@label{#1}          % use old LaTeX \label instead of new definition
                    \else
                    \label{#1}
                    \fi
                    }
\let\st@bibitem\@bibitem
\let\st@lbibitem\@lbibitem
\ifnum\draftcontrol=1
  \def\@bibitem#1{%
    \st@bibitem{#1}\a@@label{#1}\ignorespaces}
  \def\@lbibitem[#1]#2{%
    \st@lbibitem[#1]{#2}\a@@label{#2}\ignorespaces}
  \def\a@@label#1{%
    \gdef\a@lab{\smash{\normalfont\small#1}}
    \ifvmode
      \if@inlabel
        \global\setbox\@labels\hbox{%
          \llap{\a@lab\let\a@lab\relax
                \kern\@totalleftmargin\kern\marginparsep}%
          \box\@labels}%
      \fi
    \fi}
\fi
\documentclass[12pt,letterpaper]{article}
\usepackage{amsmath,bm,amsfonts,amssymb,array,calc,amsthm,rotating}
\usepackage{epsfig}
\usepackage[nosort]{cite}
\usepackage{graphicx}
\usepackage{color}
\usepackage[colorlinks=false]{hyperref}
\renewcommand\baselinestretch{1.25}
\renewcommand\section{\@startsection {section}{1}{\z@}%
                                   {-3.5ex \@plus -1ex \@minus -.2ex}%
                                   {2.3ex \@plus.2ex}%
                                   {\normalfont\large\bfseries}}
\renewcommand\subsection{\@startsection{subsection}{2}{\z@}%
                                   {-3.25ex\@plus -1ex \@minus -.2ex}%
                                   {1.5ex \@plus .2ex}%
                                   {\normalfont\normalsize\bfseries}}
\renewcommand\subsubsection{\@startsection{subsubsection}{3}{\z@}%
                                   {-3.25ex\@plus -1ex \@minus -.2ex}%
                                   {1.5ex \@plus .2ex}%
                                   {\normalfont\normalsize\it}}
\renewcommand\paragraph{\@startsection{paragraph}{4}{\z@}%
                                   {-3.25ex\@plus -1ex \@minus -.2ex}%
                                   {1.5ex \@plus .2ex}%
                                   {\normalfont\normalsize\bf}}
\renewcommand\subparagraph{\@startsection{subparagraph}{5}{\z@}%
                                   {-1.25ex\@plus -1ex \@minus -.2ex}%
                                   {0ex \@plus .2ex}%
                                   {\normalfont\normalsize\it}}

\numberwithin{equation}{section}

\long\def\@makecaption#1#2{%
  \vskip\abovecaptionskip
  \sbox\@tempboxa{{\bf #1:} #2}%
  \ifdim \wd\@tempboxa >\hsize
    {\small\bf #1:} {\small #2}\par
  \else
    \global \@minipagefalse
    \hb@xt@\hsize{\hfil\box\@tempboxa\hfil}%
  \fi
  \vskip\belowcaptionskip}
\renewcommand*\l@section[2]{%
  \ifnum \c@tocdepth >\z@
    \addpenalty\@secpenalty
    \addvspace{.5em \@plus\p@}%
    \setlength\@tempdima{1.5em}%
    \begingroup
      \parindent \z@ \rightskip \@pnumwidth
      \parfillskip -\@pnumwidth
      \leavevmode \bfseries
      \advance\leftskip\@tempdima
      \hskip -\leftskip
      #1\nobreak\hfil \nobreak\hb@xt@\@pnumwidth{\hss #2}\par
    \endgroup
  \fi}
\renewcommand*\l@subsection{\addvspace{.0em \@plus\p@}\@dottedtocline{2}{1.5em}{2.3em}}
\renewcommand*\l@subsubsection{\addvspace{-.2em \@plus\p@}\@dottedtocline{3}{3.8em}{3.2em}}

\def\hepth#1{\href{http://xxx.arxiv.org/abs/hep-th/#1}{{arXiv:hep-th/#1}}}

\definecolor{refcol}{rgb}{0.2,0.2,0.8}
\definecolor{eqcol}{rgb}{.6,0,0}
\definecolor{purple}{cmyk}{0,1,0,0}

\gdef\@citecolor{refcol}
\gdef\@linkcolor{eqcol}
\def\colorlinkspurple{\gdef\@urlcolor{purple}}
\def\colorlinksblue{\gdef\@urlcolor{blue}}
\def\colorlinksred{\gdef\@urlcolor{red}}

\def\ie{{\it i.e.}}
\def\eg{{\it e.g.}}

\def\revise#1       {\raisebox{-0em}{\rule{3pt}{1em}}%
                     \marginpar{\raisebox{.5em}{\vrule width3pt\
                     \vrule width0pt height 0pt depth0.5em
                     \hbox to 0cm{\hspace{0cm}{%
                     \parbox[t]{4em}{\raggedright\footnotesize{#1}}}\hss}}}}

\def\calh         {{\cal H}}

\def\caln         {{\cal N}}

\def\complex      {{\mathbb C}}

\def\projective   {{\mathbb P}}

\def\reals        {{\mathbb R}}
\def\zet          {{\mathbb Z}}

\def\ee           {{\it e}}
\def\ii           {{\it i}}

\def\tr           {{\rm Tr}}

\def\id           {{\rm id}}

\def\sqr#1#2{{\vcenter{\vbox{\hrule height.#2pt
 \hbox{\vrule width.#2pt height#1pt \kern#1pt
 \vrule width.#2pt}\hrule height.#2pt}}}}

\def\om{\omega}

\def\ff#1{\begin{pmatrix} #1 \end{pmatrix}}

\def\str{{\rm Str}}

\catcode`\@=12

\begin{document}

%%%
%%%%%% text starts here
%%%%%%%%%

\title{Moduli Stabilization in Non-Geometric Backgrounds}

\pubnum{%
hep-th/0611001 \\
HUTP-06/A044}
\date{October 2006}

\author{
Katrin Becker$^{a}$, Melanie Becker$^{a}$, Cumrun Vafa$^{b}$,
and Johannes Walcher$^{c}$ \ \\[0.4cm]
\it $^a$ Texas A \& M University, College Station, TX 77843, USA
\\[0.2cm]
\it $^b$ Harvard University, Cambridge, MA 02138, USA \\ and \\
\it Center for Theoretical Physics, MIT, Cambridge, MA 02139,
USA\\[.2cm]
\it $^c$ Institute for Advanced Study,
\it Princeton, NJ 08540, USA}

\Abstract{Type II orientifolds based on Landau-Ginzburg models are
used to describe moduli stabilization for flux compactifications of
type II theories from the world-sheet CFT point of view. We show
that for certain types of type IIB orientifolds which have no
K\"ahler moduli and are therefore intrinsically non-geometric, all
moduli can be {\it explicitly} stabilized in terms of fluxes.
The resulting four-dimensional theories can describe Minkowski as
well as Anti-de-Sitter vacua.  This construction provides the first
string vacuum with all moduli frozen and leading to a 4D Minkowski
background.}

%\enlargethispage{1.5cm}

\makepapertitle

\body

\version\versionno

\vskip 1em

%\tableofcontents

%\newpage

%\begin{center}
%{\bf\large Charges of O-plane and Fluxes in LG}
%\end{center}

\medskip

\section{Introduction}

With the discovery of Calabi-Yau compactifications more than twenty
years ago it became evident that many aspects of the 4D theory can
be traced back to the topology of the internal manifold. It did not
take long until backgrounds resembling the real world
were constructed. At the
same time it became evident that string theory does not have a
unique ground state because the values of the moduli fields
describing the deformations of the internal manifold could not be
determined. This has been an open problem for many years, not only
for particle phenomenology predictions coming from string theory,
but also for string cosmology. This situation changed over the past
years, as it has been realized that flux compactifications of string
theory can stabilize all the moduli fields.

Due to the incorporation of fluxes, the continuous choice of
moduli parameters was restricted to a large number of discrete
choices.  Thus this still left an extremely large number of string vacua.
These vacua are part of the string theory landscape, which at present
is analyzed with statistical methods \cite{Douglas:2003um} and techniques
borrowed from number theory \cite{DeWolfe:2004ns}.  See \cite{dk:2006}
for a review.

All the more it is surprising that the number of explicit models
known in the literature with all geometric moduli stabilized is
rather limited \cite{stabilized} and no models leading to
four-dimensional Minkowski space have been explicitly constructed.
So far, moduli stabilization has been discussed in the literature in the
supergravity approximation, a limit for which the radial modulus is
assumed to be large and the string coupling is small.
In many cases the radial modulus is then fixed in terms of
non-perturbative corrections to the superpotential in a KKLT
\cite{Kachru:2003aw} like fashion, leading to supersymmetric
Anti-de-Sitter vacua. More recently, moduli stabilization in terms
of fluxes only (i.e. at the classical level) was achieved in
\cite{villa} and in \cite{DeWolfe} in the context of type IIA
massive supergravity, where it was shown that fluxes stabilize
all geometric moduli of a simple $T^6/\zet_3\times\zet_3$
orientifold. In these models, the restrictions on the fluxes
again result in a negative cosmological constant.

One of the goals of this
paper is to construct a set of simple models in which all moduli are
{\it explicitly} stabilized by fluxes only and which have a vanishing 4D
cosmological
constant. We will do so in the context of the type IIB theory, which
allows the most freedom for dialing the fluxes. It is known that in
this theory fluxes generate a classical superpotential for the
complex structure moduli.  This is in contrast to the potential for
K\"ahler moduli, which is typically generated through non-perturbative
effects, which are less under control.

To avoid the complication of stabilizing K\"ahler moduli, the basic
idea underlying our work is to start in the type IIB theory with a
model which does not have any K\"ahler moduli to begin with, and
stabilize the complex structure moduli and the dilaton by turning on
appropriate R-R and NS-NS fluxes. With ten-dimensional supergravity in
mind, it appears quite hopeless to make any progress with this idea.
Indeed, in any {\it geometric} compactification with an ordinary
manifold $M^6$ as internal space, the overall size of that manifold
will appear as a free parameter, a K\"ahler modulus. Thus, we will
need $M^6$ to be {\it non-geometric} in one way or another. Thanks
to string theory, we know that such non-geometric models do exist.
It might be expected that the flux superpotential stabilizes all
moduli in such compactifications and that the resulting supersymmetric
vacua can be either Minkowski or AdS.

The fact that understanding string compactifications requires
generalized notions of geometry is well-appreciated. The best-known
example is probably the correspondence between sigma models on Calabi-Yau
manifolds and an effective Landau-Ginzburg (LG) orbifold model as the
``analytic continuation to small volume'' of the sigma model \cite{gvw,
wittenphases,mart:1989}. This correspondence also plays a fundamental
role in the understanding of mirror symmetry \cite{hova}.

The existence of dualities has accentuated the relevance of string
vacua without a ten-dimensional geometric interpretation. For
instance, there are examples of Calabi-Yau manifolds whose complex
structure cannot be deformed. Mirror symmetry exchanges the complex
structure with the K\"ahler structure. Therefore, the mirror duals
of such rigid manifolds would not have K\"ahler moduli and cannot
correspond to a geometric manifold. Nevertheless, they have an
effective world-sheet description as LG models, in accord with the
general ideas of \cite{hova}.

When turning on fluxes, mirror symmetry as well as other dualities
require an even broader enlargement of the allowed class of
compactification spaces. From the study of simple local or toroidal
models, it is well-known that the mirror or T-duals of
compactifications with generic fluxes cannot be described by a
conventional geometry, see \eg, \cite{schulz}. More generally, by
looking at the panoply of R-R and NS-NS fluxes that are available
in supergravity, and invoking (perturbative and non-perturbative)
dualities, one can argue that the most general flux compactification
will not allow a geometric description in any duality frame \cite{hmw}.
As mentioned above, the usual ten-dimensional effective supergravity
of string theory will not be useful for the study of such vacua.
Approaches which have been taken in the past include effective
supergravity descriptions in dimensions less than ten dimensions
\cite{hmw,hull,shelton}, as well as exact world-sheet descriptions
\cite{hmw,fww,hewa}.

In the present paper, we will use a combination of ``non-geometric''
world-sheet techniques and 4D effective space-time description to
exhibit a simple class of models in which all moduli can be
stabilized by fluxes. Depending on the particular model, different
values of the cosmological constant (Minkowski or AdS) are obtained.
Charge conservation is accomplished by the presence of orientifolds.

We will illustrate such a generic claim in a precise manner, by
studying two explicit models with Hodge numbers $h_{11}(M)=0$ and
$h_{21}(M)=84$, $90$ respectively. The underlying models before turning
on the fluxes are mirror duals of rigid Calabi-Yau manifolds and
admit an effective description as LG models \cite{vafalg}. At a
particular point in moduli space, they are also equivalent to some
Gepner model \cite{gepner}.

Our models do not have a manifold interpretation, and therefore
geometrical notions such as cycles, differential forms, etc., do not
have the conventional meaning. An appropriate description of
D-branes and supersymmetric cycles in LG models was developed in
\cite{hiv} using world-sheet techniques. A great deal of information
about D-branes and orientifolds in Gepner models and their relation
to LG and geometry is also available from the literature
\cite{GRS,bdlr,BH2,bhhw,howa,Misra:2003xx}. For convenience we
summarize the most important results in section \ref{lgmodels}.

The first new aspect of our work is the description of fluxes in
these models, which is done in section \ref{fluxes}. The flux
configurations we are interested in satisfy constraints coming from
supersymmetry and the type IIB tadpole cancellation condition.
Supersymmetric type IIB vacua have fluxes belonging to the
$H^{2,1}(M) \oplus H^{0,3}(M)$ cohomology of the internal space $M$.
We will argue that these vacua are stable even non-perturbatively.
This is due to the existence of a non-renormalization theorem for
the superpotential of \cite{Gukov:1999ya}. This basically follows
because there are no relevant instantons to correct it. The
existence of this non-renormalization theorem was crucial for the
works \cite{civ,dv} where the classically generated superpotential
(which is holographically dual to the sum of the planar diagrams of
the gauge theory) does not receive any corrections.  The fact that
these results are in complete agreement with the exact
non-perturbative dynamics of supersymmetric theories is a powerful
check on the non-renormalization of the superpotential. We will also
discuss in section \ref{fluxes} the other known consistency
conditions of type IIB flux compactifications, including flux
quantization and the tadpole cancellation condition.

In section \ref{solutions}, we will explicitly solve the
supersymmetry equations which follow from the flux superpotential
and the tadpole cancellation condition for the simpler of our two
models, related to the so-called $1^9$ Gepner model. We find
supersymmetric Minkowski vacua for several types of flux configurations.

It turns out that the spectrum of possibilities in the $1^9$ model is
quite constrained, and we have not been able to find solutions leading to
4D AdS space in this model. One might ask whether this has any significance
or whether it is just an accident in this particular case. We address this
question in section \ref{other} by repeating the analysis for the so-called
$2^6$ Gepner model. We will see that the range of possibilities is much
larger, and that we can in particular find 4D AdS solutions.

We would like to point out that we have not been able to find solutions or
sequences of solutions in which the dilaton is stabilized at very small
values, although we have no general argument why this cannot be done.
Because of the non-renormalization theorem for the
superpotential which we have mentioned above, having a string coupling of
$O(1)$ does not affect the {\it existence} of the solutions. Nevertheless,
it does mean that for other aspects of the solution, such as the masses
of the moduli, as well as introduction of supersymmetry breaking effects
which do receive quantum corrections, one should try to find a sequence
of vacua which stabilize the coupling constant at weaker values.

In the appendix, we present an analysis of type IIB orientifolds of
the quintic threefold from the LG point of view. In particular, we
discuss the computation of the O-plane charge in the LG model, as
well as the LG representation of the D3-brane sitting at a point of
the quintic. The orientifold actions which we study include
exchanges of variables, and have not been treated before in either
the Gepner model or LG literature. This discussion, which is an
application of the general setup of \cite{howa}, serves as
background for our discussion of tadpole cancellation condition in
section \ref{tadpolecanc}.

We discuss the implications of our results for studies of the string
theory landscape in section \ref{conclusions}.

\section{Branes and Orientifolds in Landau-Ginzburg models}
\label{lgmodels}

Due to their simplicity, LG models of string compactifications have
been studied in great detail over the years. They were among the
first examples in which to access ``stringy geometry''. Moreover, as
studied in \cite{gvw,wittenphases,mart:1989}, LG models can often be
thought of as the analytical continuation of Calabi-Yau sigma models
to substringy volume.

The models for which this works most straightforwardly are
Calabi-Yau hypersurfaces in weighted projective space. Namely, they
are given by the vanishing locus of a homogeneous polynomial
$P(x_1,\ldots,x_5)=0$ in five variables of weight $w_1,\ldots,w_5$,
and of total degree $H=\sum w_i$. Under the CY/LG correspondence,
this sigma model corresponds to an $\caln=(2,2)$ LG orbifold model
with five chiral fields and world-sheet superpotential $W=P$. The
orbifold group is a $\zet_H$ acting by phase rotations $x_i\to
\ee^{2\pi\ii/h_i} x_i$, where $h_i=H/w_i$, which are assumed to be
integer. For a special choice of polynomial $P$, this LG model flows
in the IR to a particular CFT with a rational chiral algebra known
as a Gepner model. These Gepner models are formally defined as the
tensor product of some number $r$ of $\caln=2$ minimal models of
level $k_i=h_i-2$, such that the total central charge is $\hat c =
\sum (1-2/h_i)=3$. Gepner models with a hypersurface interpretation
have $r=5$, in which case the central charge condition is equivalent
to the CY condition on the degree of $P$.

The correspondence between LG models and Calabi-Yau sigma models can
be extended to the boundary sector, D-branes, and orientifolds.
Boundary states in Gepner models were first constructed in
\cite{GRS}, and their geometric interpretation was first addressed
in \cite{bdlr}. In \cite{hiv}, A-branes in LG models were shown to
correspond to a particular type of non-compact cycle in the
$x$-space on which the superpotential $W$ has a constant phase. More
recently, B-branes in LG models were studied in terms of matrix
factorizations \cite{mafa}. For the extension to orientifolds, we
refer to \cite{BH2,bhhw,howa}.

In the present paper, we will be concerned with a slightly different
type of LG models which do not have a manifold interpretation, as we
have mentioned in the introduction. Nevertheless, the techniques
which are used to study LG models with such a connection can still
be applied. Let us now proceed to introduce this technology
concretely in the relevant examples, referring to the literature for
the more general discussion.

\subsection{The \texorpdfstring{$1^9/\zet_3$}{19/Z3} Gepner Model}

\subsubsection{Landau-Ginzburg model}

 Our first example is a LG model $M$ based on nine minimal models
and world-sheet superpotential
$$
W = \sum_{i=1}^9 x_i^3\,,
$$
divided by a $\zet_3$ generated by
\begin{equation}
\eqlabel{g} g: x_i\to \om x_i,\quad i=1,2,\ldots 9 \qquad
\text{and}\qquad \om \equiv \ee^{2\pi\ii/3}\,.
\end{equation}
It is easy to compute the Hodge numbers of this model. As far as
complex structure deformations are concerned, they all come from
deformations of $W$ and a basis is given by the $\zet_3$-invariant
monomials of the chiral ring $\complex[x_1,\ldots,
x_9]/(x_1^2,\ldots,x_9^2)$. In other words they are given by the
polynomials in the chiral fields
\begin{equation}
x_ix_jx_k\qquad {\rm with } \qquad i\neq j\neq k\neq i\,,
\end{equation}
and there are $h_{2,1}(M)=84$ of them. To see that there are no
K\"ahler structure deformations, we recall from \cite{vafalg} that
ground states corresponding to the even cohomology always arise from
the twisted sectors in LG orbifolds. In a $\zet_3$ orbifold, there
are only two non-trivial twisted sectors, and the first must
contribute to $h_{00}(M)=1$, while the second contributes in the
conjugate $h_{33}(M)=1$. Hence $h_{11}(M)=h_{22}(M)=0$. This
reasoning also readily implies that there are no contributions to
$h_{21}(M)$ from the twisted sectors. As a result the Hodge diamond
is
\begin{equation}
\begin{array}{cc}
      &1                  \\
   &0~~~~~~0        \\
&0~~~~~~0~~~~~~0    \\
&1~~~~~84~~~~~84~~~~~1\\
&0~~~~~~0~~~~~~0\\
&0~~~~~~0\\
&1\\
\end{array}\,.
\end{equation}

\subsubsection{Geometric description}

We can use the LG orbifold technology of \cite{vafalg} and the
above-mentioned correspondence with geometry to give alternative
descriptions of the background. To illustrate this idea consider the
simpler example of the LG model based on the quotient of
\begin{equation}
\eqlabel{bx} W= x_1^3+x_2^3+x_3^3\,,
\end{equation}
by a $\zet_3$ action which sends $(x_1,x_2,x_3) \to
\om(x_1,x_2,x_3)$. This corresponds to a $T^2$ torus model at the
$\zet_3$ symmetric point in both complex structure and K\"ahler
moduli space where
\begin{equation}
\tau=\om \qquad {\rm and } \qquad \rho= \om\,.
\end{equation}
This model has three $\zet_3$ symmetries that will be relevant for
us. Two of them act geometrically by phase rotations on the $x_i$'s,
modulo the diagonal phase rotation which we have already divided
out. These $\zet_3$'s correspond to geometric symmetries of the
$T^2$. The third, somewhat less familiar, $\zet_3$ is a so-called
``quantum symmetry'' \cite{vafalg}, and is formally identified with
the dual of the original $\zet_3$ orbifold group defining our LG
model,
\begin{equation}
\zet_3^{\rm quantum} = (\zet_3^{\rm LG})^* \cong\zet_3\,.
\end{equation}
More concretely, the quantum symmetry multiplies a state in the
$l$-th twisted sector by $\om^l$. Dividing out by $\zet_3^{\rm
quantum}$ gives back the unorbifolded LG model described by the
polynomial \eqref{bx}.

The model $M$ of our interest is based on nine cubics instead of
three, and it is natural to expect a relation to geometry via
$(T^2)^3=T^6$. The precise statement is that $M$ is mirror to a
certain rigid Calabi-Yau $Z$ which can be obtained as a quotient of
$T^6$ by a  $\zet_3 \times \zet_3$ action generated by
\begin{equation}
\begin{split}
\tilde g_{12}:&\; (z_1,z_2,z_3) \to (\om z_1,\om^{-1} z_2,z_3)\,, \\
\tilde g_{23}:&\; (z_1,z_2,z_3) \to (z_1,\om z_2,\om^{-1} z_3)\,.
\end{split}
\eqlabel{generators}
\end{equation}
Here $z_1,z_2,z_3$ are the complex coordinates of $T^6=(T^2)^3$.
This manifold has Hodge numbers $h_{11}(Z)=84$ and $h_{21}(Z)=0$.
Note that in order for $\tilde g_{12},\tilde g_{23}$ to be
symmetries, we have to fix the complex structure of the torus to be
diagonal and equal to $\tau_i=\om$ for $i=1,2,3$. Being mirror to
$Z$, $M$ can also be described as a toroidal orbifold, except that
the orbifold group does not act geometrically. To explain this in
more detail, we state that $M$ can be obtained by starting from
$T^6$, where now we fix the K\"ahler structure of $T^6$  at the
$\zet_3$ orbifold point in each $T^2$ factor, $\rho_i=\om$ for
$i=1,2,3$, and divide out by a $\zet_3\times \zet_3$ subgroup of the
$(\zet_3)^3$ quantum symmetry group. While preserving supersymmetry,
this orbifold projects out all K\"ahler moduli, so that we end up
with $h_{11}(M)=0$. The complex structure moduli of the
torus are projected, and new ones appear in the twisted sector,
for a total of $h_{21}(M)=84$.

Alternatively, we can start with the $1^9/\zet_3$ Gepner model and
turn it into a $T^6$ at the orbifold point in K\"ahler moduli space
by modding out by a $\zet_3\times\zet_3$ generated by
\begin{equation}
\begin{split}
g_{1}:&\; (x_1,x_2,x_3,x_4,x_5,x_6,x_7,x_8,x_9)\to
(\om x_1,\om x_2,\om x_3, x_4,x_5,x_6,x_7,x_8,x_9)\,, \\
g_{2}:&\; (x_1,x_2,x_3,x_4,x_5,x_6,x_7,x_8,x_9)\to (x_1,x_2,x_3, \om
x_4,\om x_5,\om x_6,x_7,x_8,x_9)\,.
\end{split}
\end{equation}
The quantum symmetry group $(\zet_3)^3$ of this $W/(\zet_3)^3$ LG
orbifold model of $(T^2)^3$ is generated by $g_1^*$, $g_2^*$ and
$g_3^*$, where $g_3=g g_1^{-1} g_2^{-1}$. The $\zet_3\times \zet_3$
which turns $T^6$ back into $M$ is then generated by $g_1^*
(g_2^*)^{-1}$ and $g_2^* (g_3^*)^{-1}$, which are mirror duals of
$\tilde g_{12}$ and $\tilde g_{23}$ in \eqref{generators},
respectively. Here, $g$ is the original orbifold generator in
\eqref{g}.

\subsection{Orientifolds}
\subsubsection{Involutions}

We intend to compactify type IIB string theory on an orientifold of
$M$, which results in an ${\cal N}=1$ theory in four dimensions. The
orientifold is defined by dividing out B-type world-sheet parity
$\Omega_B$ dressed with a holomorphic involution $\sigma$ such that
the square of it is in the orbifold group and such the
superpotential is invariant up to a sign \cite{BH2,howa}
\begin{equation}
\label{bi} W(\sigma x) = - W(x)\,.
\end{equation}
This last condition comes from the fact that superpotential enters
the world-sheet action as an F-term,
\begin{equation}
\int d\theta^+ d\theta^- W(x)\,,
\end{equation}
and B-type world-sheet parity exchanges the fermionic coordinates in
superspace $\theta^+ \leftrightarrow \theta^-$. The sign resulting
from this parity transformation is compensated for by different
types of involutions. The simplest involution that cancels the sign
in \eqref{bi} changes the sign of all nine bosonic coordinates
\begin{equation}
\eqlabel{trivial} \sigma_0 :  \; (x_1,x_2,\ldots, x_9) \to (-x_1
,-x_2,\ldots, -x_9)\,.
\end{equation}
Under this transformation the superpotential changes sign
$W(\sigma_0 x) = -W(x)$. Since we are already working with a
$\zet_3$ orbifold by $g$ in \eqref{g}, we also have to divide out by
the parity reversing symmetries $g\sigma_0$ and $g^2\sigma_0$. The
full orientifold group $\zet_3\ltimes \zet_2\cong \zet_6$ is
generated by $g\sigma_0$.

We can consider other orientifolds with a more non-trivial action on
the $x_i$'s. In particular, we can permute, respectively, 1, 2, 3,
or 4 pairs of $x_i$'s:
\begin{equation}
\eqlabel{sigmas}
\begin{split}
\sigma_1: \; (x_1,x_2,x_3,x_4,x_5,x_6,x_7,x_8,x_9),
\to -(x_2,x_1,x_3,x_4,x_5,x_6,x_7,x_8,x_9)\,, \\
\sigma_2: \; (x_1,x_2,x_3,x_4,x_5,x_6,x_7,x_8,x_9)
\to -(x_2,x_1,x_4,x_3,x_5,x_6,x_7,x_8,x_9)\,, \\
\sigma_3: \; (x_1,x_2,x_3,x_4,x_5,x_6,x_7,x_8,x_9)
\to -(x_2,x_1,x_4,x_3,x_6,x_5,x_7,x_8,x_9)\,, \\
\sigma_4: \; (x_1,x_2,x_3,x_4,x_5,x_6,x_7,x_8,x_9) \to
-(x_2,x_1,x_4,x_3,x_6,x_5,x_8,x_7,x_9)\,.
\end{split}
\end{equation}
One of the effects of the orientifold is to project the space of
complex structure deformations onto the subspace which is compatible
with $W(\sigma x)=-W(x)$. The number of invariant complex structure
deformations $h_{21}^+$, as well as invariant three-cycles, $b_3^+ =
2(h_{21}^++1)$ is tabulated in table \ref{inv}. They can be obtained
as follows. The 84 deformations of $M$ correspond to the monomials
$x_ix_jx_k$ with distinct $i,j,k=1,\ldots,9$. A parity $\sigma_i$
acts on these monomials, leaving $n_{\rm fix}(\sigma_i)$ of them
invariant up to the sign, and permuting the others pairwise. The
number of (anti-)invariant deformations is then given by
\begin{equation}
h_{21}^+(M^{(\sigma_i)}) = \frac{84-n_{\rm fix}(\sigma_i)}2 + n_{\rm
fix}(\sigma_i)\,.
\end{equation}
Eg, for $\sigma_1$, invariant monomials are $x_1x_2x_j$ with
$j=3,\ldots,9$ as well as $x_ix_jx_k$ with distinct $i,j,k=
3,\ldots,9$, so $n_{\rm fix}(\sigma_1) = 7 + 35=42$ and
$h_{21}^+=63$.

\begin{table}[t]
\begin{center}
\begin{tabular}{|c|l|l|}
\hline
orientifold & $h_{21}^+$ & $b_3^+$\\
\hline
$\sigma_0$ & 84 & 170\\
$\sigma_1$ & 63 & 128 \\
$\sigma_2$ & 52 & 106 \\
$\sigma_3$ & 47 & 96 \\
$\sigma_4$ & 44 & 90
\\\hline
\end{tabular}
\caption{Number of invariant complex structure deformations for
various orientifolds of $M$.} \label{inv}
\end{center}
\end{table}

\subsubsection{Orientifold planes}

One important piece of data of an orientifold in the geometric
setting is the fixed point locus of the involution that dresses
world-sheet parity. The connected components of this fixed point
locus are referred to as ``orientifold planes'' (O-planes). O-planes
carry R-R charge and NS-NS tension. The crucial role of O-planes
arises from the fact that both charge and tension can be {\it
negative}, while preserving space-time supersymmetry. In fact, in
compact models, O-planes are necessary in order to cancel the
tadpoles generated by D-branes and fluxes.

In our present setup, which is non-geometric, there is no
straightforward notion of orientifold ``plane'' as a geometric
locus. Nevertheless, since the charge of O-planes can be detected on
the world-sheet by computing a crosscap diagram with R-R field
insertion, we can still ask in a meaningful way for the R-R charge
sourced by the orientifold. We will present pertinent formulas for
these R-R charges in subsection \ref{oplanech}.

In geometric type IIB orientifolds, the mutually supersymmetric
O-planes that can occur together are either O9 and O5-planes or O7
and O3-planes. In other words, the complex codimension of the fixed
point locus of the dressing involution can be either $0\bmod 2$ or
$1\bmod 2$. For example, we can view the type I string as a type IIB
orientifold with trivial involution dressing world-sheet parity.
This corresponds to the O9/O5-case, where O5-plane charge can be
induced from the curvature of the compactification manifold.

In our model, we also have a similar distinction between two types
of orientifolds based on the space-time supersymmetry of the
crosscap state resulting from the various involutions. For instance,
we can anticipate that the canonical action $\sigma_0$ corresponds
to type I compactification on $M$. Then, because they are of ``even
codimension'' in the sense of having an even number of $-1$
eigenvalues, the orientifolds associated with $\sigma_2$, and
$\sigma_4$ are also of the O9/O5-type, and similarly $\sigma_1$ and
$\sigma_3$ correspond to O7/O3-type. The latter is the class of
orientifold models considered in the supergravity regime in
\cite{Giddings:2001yu}, and in which moduli can be stabilized by
fluxes. In this section, we will discuss all possible orientifolds
of $M$. The fact that even in the non-geometric LG model we study we
find only two distinct types of O-planes is a strong indication that
even in the non-geometric model, the intuition of the geometric case
continues to hold.

\subsection{D-branes and supersymmetric cycles}

\subsubsection{Supersymmetric cycles}

Before we can describe the fluxes and the tadpole cancellation
condition in our model, we need to review some background material
about the LG description of D-branes wrapping supersymmetric cycles.
Generally speaking, there are two types of supersymmetric cycles in
Calabi-Yau type compactifications: A-cycles are middle-dimensional
cycles represented by (special) Lagrangian submanifolds. They are
relevant for flux compactifications as the cycles supporting R-R and
NS-NS three-form fluxes. B-cycles are even-dimensional cycles
represented by holomorphic submanifolds carrying (stable)
holomorphic vector bundles. They are the cycles that can be wrapped
by O-planes and D-branes of various types to, for instance, support
standard model gauge fields, cancel tadpoles from fluxes, etc.

In the non-geometric setting, the most useful way to talk about
supersymmetric cycles is in terms of D-branes. Namely, we can
distinguish A- and B-branes by the boundary conditions on the
$\caln=(2,2)$ supercurrents on the world-sheet
\begin{equation}
\eqlabel{wssusy}
\begin{split}
\text{A-type:} & \qquad\qquad G^{\pm}_L(z) = G^{\mp}_R(\bar z) \quad
\text{at} \quad \bar z=\bar z\,, \\
\text{B-type:} & \qquad\qquad G^{\pm}_L(z) = G^{\pm}_R(\bar z) \quad
\text{at} \quad  z=\bar z\,.
\end{split}
\end{equation}
So let us review some of what is known about D-branes wrapping
supersymmetric cycles in LG models. There are generally two
languages to describe the cycles. In \cite{hiv}, A-branes in LG
models were related to Picard-Lefshetz vanishing cycles of the
singularity described by the LG superpotential. In the simple
homogeneous cases, these cycles can be pictured as wedges in the
$x$-plane, see \eg, Fig.\ \ref{piece}. It is straightforward to
implement orbifolding in this description. The other language we
will use is the more abstract language of matrix factorizations,
originally introduced in \cite{mafa}. These matrix factorizations
describe B-cycles in LG models and their orbifolds and have been
studied extensively in the last few years.

Since A-model and B-model are mirror to each other, and the mirror
of an LG model is an orbifold of it, we can also combine the roles
of these two descriptions. The wedge description is easier to
picture, while the matrix factorization approach is more general, in
the sense that not all matrix factorizations have a (known) mirror
wedge description. Moreover, the O-plane charge is in general only
known how to compute in this language.

We will first describe these cycles in the parent LG model
(including the orbifold, but before orientifolding). Later, we will
describe how the orientifold projects them and work out invariant
representatives.

\subsubsection{A-branes}
\label{Abranes}

\begin{figure}[t]
\begin{center}
\epsfig{height=6cm,file=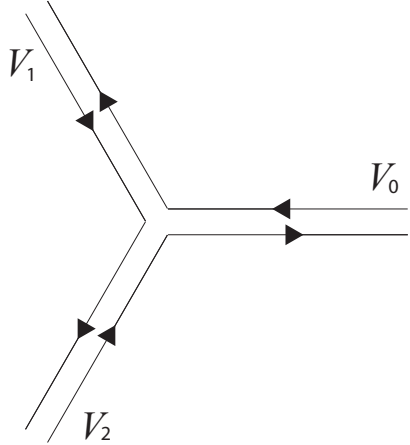} \caption{The piece of cake
picture of Lagrangian (A-type) D-branes in LG model $x^3$.}
\label{piece}
\end{center}
\end{figure}
By studying the A-type world-sheet supersymmetry condition
\eqref{wssusy}, one finds \cite{hiv} that A-type D-branes in LG
models correspond to middle-dimensional cycles. In the $x$-space
D-branes correspond to the preimages of the positive real axis (or,
by an R-rotation, some other straight ray with constant slope) in
the $W$-plane, {\it i.e.} to the preimages of
\begin{equation}
\eqlabel{Atype} {\rm Im} (W) = 0\,.
\end{equation}
In the simplest case of LG models, based on the superpotential
$W=x^{k+2}$ (and its deformations), this condition can be solved
completely, and one can compare with RCFT results on boundary states
in $\caln=2$ minimal models. Specifically, the condition
\eqref{Atype} selects $(k+2)$ spokes through the origin in the
$x$-plane at an angle which is an integer multiple of
$\ee^{2\pi\ii/(k+2)}$, and the branes correspond to all possible
wedges built from these spokes. A useful homology basis is provided
by the $(k+2)$ wedges $V_0,\ldots V_{k+1}$ of smallest angular size
$\ee^{2\pi\ii/(k+2)}$.

For example, for the minimal model building block of the $1^9$
Gepner model, each of the factors comes with a set of three
A-branes, see Fig.\ \ref{piece}. We will call these three cycles
$V_0$, $V_1$, and $V_2$. The $V_n$'s generate the homology of the
minimal model, but satisfy the one relation
\begin{equation}
\eqlabel{relation} V_0 + V_1 + V_2 = 0 \,,
\end{equation}
as can be seen from the figure.

On the other hand, the cohomology basis of the space of A-type
D-brane charges in the minimal model is spanned by the two R-R
sector ground states, $|l\rangle$, with $l=1,2$ \cite{vafalg}. These
can be equivalently represented by the chiral ring
$\complex[x]/x^2=\langle 1, x\rangle$. The correspondence is given
by
\begin{equation}
\eqlabel{corr} |l\rangle\leftrightarrow x^{l-1}\,.
\end{equation}
The R-R charges of the $V_n$'s can be computed as the overlaps
\cite{hiv} (disk one-point functions)
\begin{equation}
\eqlabel{overlap} \langle V_n|l\rangle = \int_{V_n}
x^{l-1}\ee^{-x^3} dx= (1-\om^l) \om^{ln}\,.
\end{equation}
The normalization we have chosen here differs slightly from the ones
of  \cite{hiv}, but is more convenient for our purposes.

\subsubsection{Intersection form of A-cycles}

For later applications, it is useful to discuss the action of the
$\zet_3$ symmetry, as well as the intersection form on the A-cycles
in the minimal model.

First of all, it is quite obvious that the $\zet_3$ symmetry which
sends $x\to \om x$ is represented on the $V_i$'s as
\begin{equation}
(V_0,V_1,V_2)\to(V_1,V_2,V_0)=(V_0,V_1,V_2) g\qquad {\rm with }
\qquad g=\begin{pmatrix}0 & 0 & 1\\  1 & 0 & 0 \\ 0 & 1 & 0
\end{pmatrix}\,.
\end{equation}
Turning to the intersection form, this is best defined as the Witten
index $\tr(-1)^F$ in the Hilbert space of open strings between two
branes. Geometrically, there is clearly one R-R ground state
localized at each geometric intersection point. In the more abstract
setting such as the LG models considered herein, we can still
compute $\tr(-1)^F$ as the cylinder amplitude for open strings
stretched between two branes with supersymmetric boundary conditions
around the cylinder. In the limit that the cylinder becomes very
long, this amplitude factorizes on disk amplitudes with closed
string ground states inserted, thus providing an alternative (and
often very simple) derivation of D-brane charges from the Witten
index. In the case at hand, the intersection product between $V_m$
and $V_n$, which is defined as $\tr_{\calh_{mn}}(-1)^F$, follows
most easily from the relation to soliton counting \cite{hiv}. One
finds
\begin{equation}
\eqlabel{intersect} \tr_{\calh_{mn}} (-1)^F = \id - g = \ff{1 & 0 &
-1\\ -1 & 1 & 0 \\ 0 & -1 & 1}\,.
\end{equation}
The ``index theorem'' which expresses this in terms of the R-R
overlap \eqref{overlap} is explicitly
\begin{equation}
\eqlabel{closed} \tr_{\calh_{mn}} (-1)^F = \frac13 \sum_{l=1,2}
\langle V_m|l\rangle \frac{1}{1-\om^l} \langle l|V_n\rangle\,,
\end{equation}
where $\langle l|V_n\rangle=\overline{\langle V_n|l\rangle}$.

Clearly, \eqref{intersect} does not have full rank, which is a
reflection of the relation in homology \eqref{relation}. We can
truncate the $\zet_3$ representation and intersection matrix by
passing to a basis of A-cycles, given for instance by $(V_0,V_1)$.
The $\zet_3$ generator takes the form
\begin{equation}
A = \ff{0 & -1 \\ 1 & -1}\,,
\end{equation}
while the intersection matrix is
\begin{equation}
I = \ff{1 & 0 \\ -1 & 1}\,.
\end{equation}
\def\l{{\bf l}}
\def\n{{\bf n}}
\def\m{{\bf m}}
We now tensor together nine such minimal models and orbifold by $g$
acting diagonally as in \eqref{g}. The orbifolding projects R-R
ground states and chiral ring to those states
\begin{equation}
\eqlabel{those}
|\l\rangle=|l_1,\ldots, l_9\rangle\qquad {\rm with } \qquad l_i=1,2
\qquad {\rm and } \qquad  \sum_{i=1}^9 l_i=0\bmod 3\,,
\end{equation}
and identifies brane states by summing over orbits. We will denote
these branes as
\begin{equation}
\eqlabel{correct} \Gamma_{[\bf n]} = \frac{1}{\sqrt{3}}(\otimes_{i}
V_{n_i} + \otimes_i V_{n_i+1} + \otimes_i V_{n_i+2}) \qquad {\rm
with } \qquad {\bf n}=(n_1,\ldots n_9)\,,
\end{equation}
where the $n_i$'s are taken $\bmod ~3$. To see that the factor
$1/\sqrt{3}$ on the RHS is the
correct normalization factor of the boundary states, we look at the
open string spectrum. Let us look in particular at the intersection
index $\tr(-1)^F$ between $\Gamma_{[\bf m]}$ and $\Gamma_{[\bf n]}$.
In the parent (unorbifolded) LG model, each of them has three
pre-images rotated by $g$. It is clear that the intersection of the
two branes in the orbifold is given by looking at the intersection
points of all these 9 preimages. Now if we rotate the preimages of
both branes simultaneously, the intersection does not change,
trivially because $g$ is a symmetry. In other words, the
intersection points related by rotating both preimages
simultaneously are gauge equivalent. Therefore, to obtain the
intersection in the orbifold, we fix any one preimage of
$\Gamma_{[\bf m]}$ and look for intersections with all preimages of
$\Gamma_{[\bf n]}$. Thus, the intersection form on the $\Gamma_{[\bf
n]}$ is given by \cite{bdlr}
\begin{equation}
(1-g)^{\otimes 9} \left[1 + g^{\otimes 9} + (g^{\otimes
9})^2\right]\,,
\end{equation}
or, restricted to those $\Gamma_{[\bf n]}$ with representatives with
all $n_i=0,1$,
\begin{equation}
\eqlabel{inters} {\bf I} =  I^{\otimes 9} \left[ 1 + A^{\otimes 9} +
(A^{\otimes 9})^2\right]\,.
\end{equation}
The same result can also be expressed in a form similar to \eqref{closed}
\begin{equation}
\eqlabel{similar} {\bf I}_{\n\m} = \frac 1{3^8} \sum_{\l} \prod
(1-\om^{l_i})\frac{1}{\prod (1-\om^{l_i})} \prod (1-\om^{-l_i})
\om^{\n\cdot\l-\m\cdot\l}\,,
\end{equation}
where $\n\cdot\l = \sum n_i l_i$ and the sum is over all $\l$ with
$\sum l_i=0\bmod 3$, cf.\ \eqref{those}. Since we are summing over
$170$ intermediate states in \eqref{similar}, the $2^9\times
2^9$-dimensional matrix ${\bf I}$ has rank $\le 170$.
As it turns out, when restricting ${\bf I}$ to the first
$170$ (in alphabetical order) of the $\Gamma_{[\bf n]}$ with
$n_i=0,1$, ${\bf I}$ is invertible and has determinant $1$. We have
thus described an algorithm for finding a minimal integral basis of
A-cycles in our $1^9$ LG model.

\subsubsection{B-branes and matrix factorization}
\label{Bbranes}

The analog of \eqref{Atype} for B-type boundary condition
\eqref{wssusy} would naively impose the holomorphic condition $W=0$
at the boundary. Both conditions arise from the fact that the F-term
world-sheet interaction $\int d^2\theta W$ is supersymmetric only up
to partial integration, and picks up a contribution from the
boundary if one is present. In the case of A-branes, this boundary
term is eliminated by imposing the boundary condition \eqref{Atype}.
But for B-branes, restricting to $W=0$ would not allow for a great
diversity of boundary conditions. In that case one introduces
additional boundary degrees of freedom and boundary interactions
whose susy variation will cancel the boundary term.

As it was shown in \cite{mafa}, one way of encoding these boundary
interactions is in terms of {\it matrix factorization}. Briefly, a
matrix factorization of the world-sheet superpotential $W$ is a
block off-diagonal matrix $Q$ with polynomial entries in the LG
variables and satisfying the equation
\begin{equation}
Q^2 = W\cdot{\rm id}\,.
\end{equation}
Physically, the boundary interactions encoded by $Q$ can be viewed
as an open string tachyon configuration between space filling branes
and anti-branes: The blocks of $Q$ correspond to the Chan-Paton
spaces of the branes, resp., the anti-branes. The off-diagonal blocks
correspond to the open string tachyon. The diagonal blocks
could carry a gauge field configuration, which however can be gauge
away in the standard LG models.

For example, for the minimal model $x^3$, there is essentially only
one non-trivial factorization $x^3 = x\cdot x^2$, with associated
matrix factorization
\begin{equation}
\eqlabel{Q} Q = \ff{0 & x\\ x^2 & 0} \qquad Q^2 = x^3 \cdot \id\,.
\end{equation}
The spectrum of massless open strings between two such branes is
computed as the cohomology of the matrix factorizations acting on
matrices with polynomial entries. For the simple example \eqref{Q}
for instance, it is easy to see that there are exactly two such
states,
\begin{equation}
\bigl[Q, \ff{1&0\\0&1}\bigr] = 0, \qquad
\bigl\{Q,\ff{0&1\\-x&0}\bigr\} =0\,.
\end{equation}
We must refer to the original literature \cite{mafa}, and citations
thereof, for more details about the matrix factorization description
of B-branes in minimal models. We will review the essentials here in
view of their applications in our models.

Now let us discuss B-branes in orbifolds. The simplest, and also
useful, example is the $\zet_3$ orbifold of a single minimal model
$x^3$ divided by the action $x\to\om x$. As we have mentioned, this
orbifolding is nothing but an implementation of mirror symmetry, so
we should compare the result with the wedge picture of A-branes in
the (unorbifolded) model we have described above. Since we are
dealing with space filling branes, we seek a representation of the
orbifold group on their Chan-Paton spaces. It is easy to see that
$\zet_3$-generator is represented on the matrix $Q$ of \eqref{Q} by
\begin{equation}
\eqlabel{gamma} \gamma = \om^n \ff{1 & 0 \\ 0 & \om}\,,
\end{equation}
where $n=0,1,2$. This generator satisfies
\begin{equation}
\gamma Q(\om x) \gamma^{-1} = Q(x)\,.
\end{equation}
The factorization $Q$ equipped with these three representations of
$\zet_3$ corresponds via mirror symmetry of the minimal model with
its orbifold precisely to the three A-type wedges $V_n$ discussed in
the previous subsection. One can also work out the projection on the
open strings and thereby recover \eqref{intersect}. The expressions
\eqref{overlap} and \eqref{closed} are then special cases of the
general formulas in \cite{stability}.

For our purposes, we are interested in the orbifold of
$W=\sum_{i=1}^9 x_i^3$ by a diagonal $\zet_3$. Before orbifolding,
we have just one factorization, given by tensoring nine copies of
$Q$ in \eqref{Q}. In the orbifold, this yields three different
branes: We tensor together nine copies of $\gamma$ in \eqref{gamma},
with naively $3^9$ choices of representation, but clearly only the
sum of $n$'s ($\bmod 3$) matters. We'll now call these three branes
$\Lambda_n$, $n=0,1,2$.  A different way to think of the $\Lambda_n$'s is
as A-branes in the mirror $W/\zet_3^8$, where we are dividing by all
symmetries that leave the product of $x_i$'s invariant. This
orbifold action allows to ``align'' the wedges in all nine factors,
so $\Lambda_n$ can be identified with $V_n$ in one of the factors,
\eg, the first one.

\subsubsection{Minimal integral basis for B-cycles}

{} From either one representation, we find that the intersection
matrix of the $\Lambda_n$'s is given by
\begin{equation}
{\bf J} = \ff{ 0 & -81 & 81 \\ 81 & 0 & -81 \\ -81 & 81 & 0}\,.
\end{equation}
For example, the evaluation of the formulas in \cite{stability} reads
\begin{equation}
\tr_{\calh_{\Lambda_m,\Lambda_n}} (-1)^F = \frac 13 \sum_{k=1,2}
\om^{km}(1-\om^k)^9 \frac{1}{(1-\om^k)^9} \om^{-kn}(1-\om^{-k})^9\,.
\end{equation}
{}From this, we could also read off the overlaps of the B-type
boundary states with Ramond ground states $|k\rangle$ in the twisted
sectors $k=1,2$,
\begin{equation}
\eqlabel{eg} \langle\Lambda_n| k\rangle = \str g^k = (1-\om^k)^9
\om^{k n}\,,
\end{equation}
where $g=\gamma^{\otimes 9}$, {\it cf.}, \eqref{overlap}.

In distinction to the A-brane situation, we see that the
$\Lambda_n$'s do not contain a minimal integral basis of the B-type
charge lattice. For the later discussion of tadpole cancellation in
the flux models, we will however need to know the precise
quantization condition, so it is necessary to digress a little
further on the construction of such a minimal basis.

In the context of matrix factorizations, it is by now well-known how
to construct the minimal basis. Namely, one has to use
factorizations which are not obtained as tensor products of minimal
model factorizations. Eg, the potential $x_1^3+x_2^3$ has the
factorization\footnote{These factorizations are also known as
``permutation branes'' \cite{permutation}. The mirror A-model
description of these branes is not known in the LG formulation.}
\begin{equation}
\eqlabel{perms} Q^{(12)}= \ff{0 & x_1+x_2\\ x_1^2-x_1 x_2+x_2^2 &
0}\,,
\end{equation}
which is not the tensor product of minimal model Cardy states. Such
a tensor product would be a $4\times 4$ matrix. It is thus no
surprise that $Q^{(12)}$ is also ``smaller'' as far as R-R charges
are concerned. For instance, the diagonal $\zet_3$ is  represented on
$Q^{(12)}$ by a single copy of $\gamma$ in \eqref{gamma} instead of
two. Thus, if we denote by $\Lambda^{(12)}_n$ the branes obtained by
tensoring together the $Q^{(12)}$ in \eqref{perms} with $7$ copies
of the Cardy brane \eqref{Q}, their charges are
\begin{equation}
\eqlabel{also} \langle\Lambda^{(12)}_n|k\rangle = (1-\om^k)^8
\om^{kn}\,,
\end{equation}
and their mutual intersections are
\begin{equation}
\tr_{\calh_{\Lambda^{(12)}_m,\Lambda^{(12)}_n}}(-1)^F = \ff{0 & -27
& 27 \\ 27 & 0 & -27 \\ -27 & 27 &0}\,.
\end{equation}
This is still not minimal, but it's clear how to proceed. We denote
\eg, by $\Lambda^{(12)(34)}$ the tensor product of $Q^{(12)}$ with
$Q^{(34)}$ and 5 copies of $Q$, and then with obvious further
notation, we find the overlaps
\begin{equation}
\eqlabel{also2}
\begin{split}
\langle\Lambda^{(12)(34)}_n|k\rangle &= (1-\om^k)^7 \om^{kn}\,, \\
\langle\Lambda^{(12)(34)(56)}_n|k\rangle &= (1-\om^k)^6 \om^{kn}\,, \\
\langle\Lambda^{(12)(34)(56)(78)}_n|k\rangle &= (1-\om^k)^5
\om^{kn}\,.
\end{split}
\end{equation}
The intersection matrix of the last set is $\left(
\begin{smallmatrix}0&-1&1\\1&0&-1\\-1&1&0\end{smallmatrix}\right)$,
yielding a minimal basis. It is also not hard to express the charges
of these branes built from \eqref{perms} in terms of the standard
$\Lambda_n$'s. Using $(1-\om)^{-1}=(1-\om^2)/3$, we find by comparing
\eqref{eg} with \eqref{also}
\begin{equation}
\eqlabel{permcharges}
\begin{split}
[\Lambda^{(12)}_n] &= \frac{[\Lambda_{n}] - [\Lambda_{n+2}]}3\,, \\
[\Lambda^{(12)(34)}_n] &= - \frac{[\Lambda_{n+2}]} 3\,,\\
[\Lambda^{(12)(34)(56)}_n] &= -\frac{[\Lambda_{n+2}] - [\Lambda_{n+1}]}9\,, \\
[\Lambda^{(12)(34)(56)(78)}_n] &=  \frac{[\Lambda_{n+1}]} 9\,.
\end{split}
\end{equation}

\subsection{Charges of O-planes}
\label{oplanech}

The charges of the O-planes\footnote{As we have mentioned before,
rather than implying that there is a geometric locus which can be
identified as an ``O-plane'', we here simply mean the abstract
world-sheet concept.} associated with the orientifold actions
$\sigma_i$ described in \eqref{sigmas} can be computed using the
formula (5.37) in \cite{howa}. This formula says that, in the same
basis of charges in which D-brane charges are given by the formulas
in \cite{stability}, such as we have, \eg, used them in \eqref{eg},
\eqref{also}, \eqref{also2},
the charge of the O-plane, namely the overlap of the crosscap state
$|C\rangle$ with the Ramond ground state $|k\rangle$, is given by%
\footnote{This formula gives the contribution to the charge from the
internal theory only. The spacetime contribution is universal and
multiplies the formulas in \cite{howa} by $4$. See the next section
for further discussion.}
\begin{equation}
\langle C | k \rangle = \prod_{i=1}^9 (1+ \sigma_i)\,,
\end{equation}
where $\sigma_i$ are the eigenvalues of the element of the
orientifold group which squares to $g^k$ (where $g$ is the generator
of the orbifold group).

For example, let us consider the ``trivial'' orientifold action,
\eqref{trivial}. The orientifold group has three elements which
reverse world-sheet parity, namely $\sigma_0$, $g\sigma_0$ and
$g^2\sigma_0$. To determine $\langle C|k\rangle$ with $k=1$, we
notice that $(g^2\sigma_0)^2=g$ and the eigenvalues of $g^2\sigma_0$
are $(-\om^2,\ldots,-\om^2)$ for $i=1,\ldots 9$. Thus, $\langle
C|1\rangle = (1-\om^2)^9 = - (1-\om)^9=-\langle\Lambda_0|1\rangle$.
Similarly, $\langle C|2\rangle=-(1-\om^2)^9=-\langle
\Lambda_0|2\rangle$.

Next, we consider the orientifold action involving a single
permutation of variables. The eigenvalues of $g^2\sigma_1$ are
$(\om^2,-\om^2,\ldots, -\om^2)$, and $\langle C|1\rangle=
(1+\om^2)(1-\om^2)^8=-\om^2 (1-\om)^8$. Instead of going through the
computations for the remaining cases, let us simply quote the result
for the topological classes of the O-planes $O_i$ associated with
$\sigma_i$. In terms of the basis $\Lambda_n$ ($n=0,1,2$), we find
\begin{equation}
\eqlabel{och}
\begin{split}
[O_0] &= -[\Lambda_0]\,, \\
[O_1] &= \frac{[\Lambda_1]-[\Lambda_2]}3\,, \\
[O_2] &= \frac{[\Lambda_0]}3\,, \\
[O_3] &= -\frac{[\Lambda_1]-[\Lambda_2]}9\,, \\
[O_4] &= -\frac{[\Lambda_0]}9\,. \\
\end{split}
\end{equation}
Notice that these charges coincide with the charges of particular
``permutation branes'' \eqref{permcharges}. However, this is an
accident of having level $1$ minimal models. For example, a similar
statement is not true on the quintic. The charge of the permutation
brane associated with the factorization $x_1^5+x_2^5 =
(x_1+x_2)(\cdots)$, although it owes its existence to the
permutation $x_1\to x_2$, {\it does not} coincide with the charge of
the orientifold associated with that permutation. We discuss this
explicitly in appendix \ref{quintic1}.

\section{Fluxes in Landau-Ginzburg models}
\label{fluxes}

Because our models do not have a radial modulus, it will now be
shown that all moduli of the internal space can be stabilized in
terms of fluxes. Our solution is exact, i.e. there are no
perturbative or non-perturbative corrections in the string coupling
constant. The reason is that our analysis is based on two
ingredients, the supersymmetry constraints following from the
space-time superpotential and the tadpole cancellation condition. As
argued in this section, both equations are {\it exact}.

\subsection{Flux superpotential}

Let us first discuss the situation in the geometric case, and then
explain how it continues to hold in the non-geometric LG case of interest
to us. In the type IIB theory there are three-form fluxes in the R-R
and NS-NS sector, $H_{RR}$ and $H_{NS}$ respectively, that can be
combined into a complex three-form
\begin{equation}
G=H_{RR}-\tau H_{NS}\,.
\end{equation}
Here $\tau=C_0+ie^{-\phi}$ is the axion-dilaton combination. In the
type IIB theory the fluxes generate a space-time superpotential for
the complex structure moduli
\begin{equation}
W=\int_M G \wedge \Omega\,.
\end{equation}
This superpotential was derived in \cite{Gukov:1999ya} and can be
obtained with two different arguments. First, supersymmetry
constrains the form of the allowed flux components. These
constraints were derived for M-theory/F-theory on four-fold
compactifications in \cite{Becker:1996gj}, \cite{Dasgupta:1999ss}
and the superpotential is such that it reproduces these constraints.
Similarly, the type IIB superpotential can be derived from the
supersymmetry constraints imposed on the fluxes in type IIB. The
second method involves general arguments which relate the tensions
of domain walls to the superpotential \cite{Gukov:1999ya},
\cite{Gukov:1999gr}.  As we will elaborate in the next subsection this
latter derivation of the superpotential can be used to show that the
superpotential is exact and does not receive any corrections,
perturbative or non-perturbative, beyond the tree level.

Unbroken supersymmetry demands
\begin{equation}
 D_iW= {\partial}_iW+{\partial}_iKW=0\,.
\end{equation}
Here $i$ runs over the complex structure moduli and $\tau$. Further
$K$ describes the K\"ahler potentials for the complex structure
moduli and the dilaton-axion.
\begin{equation}
K=K(z^a)+K(\tau)\,,
\end{equation}
\begin{equation}
K(\tau)=-{\rm log}[-i(\tau-\bar \tau)], \quad K(z^a)=-{\rm log} (
i\int_M \Omega\wedge \bar \Omega)\,.
\end{equation}
Demanding
\begin{equation}
D_{\tau}W=\frac{-1}{(\tau-\bar \tau)}\int_M \bar G \wedge
\Omega=0,\quad  D_aW=\int_M G\wedge \chi_a=0\,,
\end{equation}
where $\chi_a$ is a basis of harmonic $(2,1)$ forms, leads to the
conclusion
\begin{equation}
G=H_{RR}-\tau H_{NS} \in H^{2,1}(M) \oplus H^{0,3}(M)\,.
\end{equation}
Since for the first case the superpotential vanishes it corresponds
to a Minkowski solution, while the second option corresponds to AdS.
We will restrict to supersymmetric vacua, so that our analysis can
be based purely on a solution to the supersymmetry constraints.
Notice that the situation here is different than for geometric
models, where a $(0,3)$ component breaks supersymmetry due to the
presence of the radial modulus.

\subsection{Non-renormalization theorem}

It is important for our subsequent analysis that this superpotential
does not receive any perturbative nor non-perturbative corrections,
neither in $\alpha'$ nor in the string coupling constant $g_s$.
Since we are dealing with type IIB model, $\alpha'$ corrections are already
summed up in the LG model.  We will thus focus on the potential $g_s$
corrections. This will guarantee that our solutions are valid to all
orders in perturbation theory and even non-perturbatively.  We will
argue that $W$ does not receive perturbative nor any non-perturbative
corrections\footnote{There are other similar arguments for the perturbative
non-renormalization. Despite the
subtlety that the space-time superpotential depends explicitly on
the dilaton, this can be shown perturbatively using the type IIB
R-invariance, Peccei-Quinn symmetry as well as $SL(2,Z)$ invariance
\cite{Burgess:2005jx}.}. The arguments for the non-renormalization of
$W$ in the geometric case are already known and used in the literature.
Here we will elaborate them in detail as it is important to argue for
its non-renormalization even in the non-geometric case which is the
case of interest for us.

First the geometric case: Consider type IIB D5-branes and suppose we
have some $H_{RR}$ flux turned on in the internal Calabi-Yau
manifold.    If we wrap D5-branes over a three-cycle in the
Calabi-Yau, and let it be a domain wall in space-time, then the flux
value jumps from one side of the domain wall to the other.  The BPS
computation for the tension of this domain wall is by definition the
change in the value of the superpotential $W$ as we go from one side
to the other.  On the other hand the BPS tension of the D5-brane
wrapped on a 3-cycle $C$ is given by\footnote{This point will be
made more precise below.}
\begin{equation}
T=\int_C \Omega\,.
\end{equation}
Since we have
\begin{equation}
\Delta W= T=\int_C \Omega\,,
\end{equation}
and $H_{RR}$ has changed by one unit over each
3-cycle that intersects $C$ in the positive sense, this implies that
\begin{equation}
W=\int_M H_{RR}\wedge \Omega\,.
\end{equation}
Similarly if we also have $H_{NS}$ and adapt the above argument to
the NS 5-brane we have a similar story (as can also be deduced by
the S-duality of type IIB) yielding
\begin{equation}
W=\int_M G\wedge \Omega\,.
\end{equation}
Thus the question of whether there are quantum corrections to this
formula translates to the question of whether there are corrections
to the BPS tension of the D5-brane domain walls.  It is known that this is not
renormalized.  To see this first of all note that by T-duality
in spacetime part (viewing the 4d on $T^4$) this is related to
the quantum correction in the tension of D3-branes.  But
 it is well known that the BPS tension of electrically charged
states is exact at the tree level (in the ${\cal N}=2$ terminology.
This follows from the fact that string coupling constant is a
hypermultiplet whereas the tension of the D3-brane is determined by
vector multiplet data, which do not interact with hypermultiplet
terms in holomorphic terms).   By S-duality this leads to the above
formula for the tension of the NS 5-brane domain walls as well and
also to its non-renormalization.

Another way to argue for non-renormalization
 is to note that the only branes that could
have corrected the tension of D5-branes, should have been Euclidean
instantons which break half the supersymmetry and have
three-dimensional world-volume\footnote{One may have also worried
about potential correction to the tension of the $D5$-brane domain
walls from the $D$-instantons.  This could not have done the job by
itself, without wrapped internal D-brane instantons, because we know
that for pure $D(-1)$ instantons, which would have therefore been
present in the decompactified limit (as it does not depend on the
internal moduli) there is no contribution due to higher
supersymmetry.}:  They would have wrapped the corresponding internal
three-cycles of the Calabi-Yau.  But, the type IIB has no Euclidean
two-brane which preserves supersymmetry. Therefore there is no
candidate instanton.

The non-renormalization
theorem is crucial for us, because we will fix the coupling
constant at a value of order $1$.
The non-renormalization theorem for the superpotential
has passed some highly non-trivial checks:  In the context
of ${\cal N}=1$ holography studied in \cite{civ,dv} this
statement was equivalent with the statement that the exact
non-perturbative superpotential fixing the glueball
vevs of the gauge theory can be computed exactly
by considering the planar diagrams only, which in turn
is equivalent to saying that the superpotential, determined
by the fluxes is exact at the tree level (which automatically
sums up the planar diagrams).  Note that the $1/N$ corrections
to the ${\cal N}=1$ superpotential would translate
directly to $g_s$ corrections and if there were such corrections
it would have ruined the exact results of \cite{civ,dv}.
 Needless to say, the fact that
this reproduces exact non-perturbative results for gauge theories
is a strong check for the validity of the non-renormalization
theorem of the superpotential.

Now we turn to the non-geometric LG case, which is the case of main
interest for us in this paper. In such cases before even considering
the superpotential we first have to argue that the corresponding
$H_{NS}$, $H_{RR}$ degrees of freedom exist!  Since the internal CFT
is not geometric, we cannot identify $H_{RR}$ and $H_{NS}$ with
three-forms in the internal theory.  However the notion of degree of
the form can be replaced by the notion of the internal $U(1)$ charge
of the underlying $(2,2)$ SCFT.  We would like to argue that for
each chiral field $\Phi$ with charge $(1,-1)$ field, which in our
notation, corresponds to the $H^{2,1}(M)$ elements, there exists
complex $H_{NS}$ and $H_{RR}$ fields (complexification corresponding
to the $H^{1,2}(M)$ elements).  In fact we can use the world-sheet
construction to write down the corresponding vertex operators which
is most naturally done in the Berkovits' hybrid formalism
\cite{Berkovits:1995cb,Linch:2006ig,Lawrence:2004zk}:
\begin{equation}
\begin{split}
(\epsilon_{ij} (q^i_Lq^j_L-q^i_Rq^j_R)) \cdot \Phi +c.c.\qquad
&\leftrightarrow H_{NS} \,, \\
\epsilon_{ij} q^i_L q^j_R \cdot \Phi +c.c. \qquad
& \leftrightarrow H_{RR}\,,
\end{split}
\end{equation}
where $q^i_{L,R}$ denote the left/right supersymmetry generators.\footnote{
In this language the turning on of the auxiliary fields in the ${\cal N}=2$
supersymmetry multiplet is what is responsible for the generation of the
superpotential.  Moreover the non-renormalization of the superpotential
would then follow directly
from the non-renormalization of the prepotential of the ${\cal N}=2$ theory.
This was in particular the description used in \cite{vafaaug}.
Basically the point is that if $\Phi$ denotes a vector multiplet
${\cal N}=2$ superfield and giving vev $v$ to its $\theta^2$ components yields
\begin{equation*}
\int d^4 \theta {\cal F}_0(\Phi)=\int d^2\theta v\cdot \partial {\cal F}_0\,,
\end{equation*}
which leads to the above formula for $W$ when we include all contributions.
This view of the non-renormalization theorem is nice in that it follows
directly from the 4d data, without assuming anything about whether
the internal theory is geometric or not.  In particular the non-renormalization
of the prepotential ${\cal F}_0$ which is crucial for the exact computations
in the context of the ${\cal N}=2$ theories directly leads to the non-renormalization
theorem for the case with fluxes.}

We can now formulate the non-renormalization of $W$ along the lines
of the arguments discussed in the geometric case. Since we have
identified all the relevant objects in terms of the internal CFT
theory, we can apply it to the LG case.  In particular the
superpotential can be viewed as such an object:  The internal
D-branes of the geometric case, fixing the superpotential in the
geometric case, can now be translated to the non-geometric case,
simply by formulating the objects in the CFT language. For example
the notion of a $D3$-brane wrapping an internal cycles clearly has a
CFT description.  This directly leads to the CFT definition for a
$D5$-brane (simply by extending the D-brane in 2 of the spatial
directions).  Moreover the lack of instantons to correct the $D3$
brane tension, still holds as in the geometric case (the ${\cal
N}=2$ BPS charges are not renormalized, as evidenced by the absence
of relevant geometric objects; also the notion of the brane tension
certainly makes sense and the jump of flux across the corresponding
domain wall can also be formulated).  Similarly the
non-renormalization of the superpotential in terms of the lack of
availability of suitable branes follows in a similar form.  We can
thus formulate all the relevant ingredients for non-renormalization
of $W$ in the non-geometric case.

There is however, one point to consider in more detail: Note that there is
no similar non-renormalization
argument for the K\"ahler potential $K$. Therefore one may worry
about the renormalization of the criticality of the potential, namely
the condition
\begin{equation}
D_iW=0
\end{equation}
also involves the K\"ahler potential $K$.  First of all note that this
issue does not exist in the case of Minkowski solutions that we will consider
because in that case $W=0$ and thus $D_iW=\partial_i W$.  However for our
AdS type solutions we would need to argue about the non-renormalization
of $D_iW$.  This may sound impossible if $K$ gets renormalized.  However
we now argue that this is indeed the case.

First of all note that $W$ is not a holomorphic {\it function} but a
holomorphic {\it section} of a line bundle.  The fact that instead
of $\partial_i W$ we have the covariantized $D_iW$ reflects this
fact. In particular when we mentioned that the tension of the domain
wall does not get renormalized and wrote the tension as an integral
of the holomorphic three-form $\Omega$, this reflects the fact that
$W$ is a section of a line bundle.  In this case the line bundle
corresponds to the rescaling of
\begin{equation}
\Omega \rightarrow \lambda \Omega \,.
\end{equation}
Thus the worry would have been if the covariantization of derivative
could receive quantum corrections.
The solution we have found for the flux extremization states
that the flux $G$ should lie in the $H^{0,3}\oplus H^{2,1}$.  Since the rescaling
of $\Omega$ does not affect this statement, even if the section
receives quantum corrections, and may affect
how $W$ is expressed as a function, it would not affect the
form of the solution we have found which is gauge invariant, i.e.
invariant under the rescaling of $\Omega$.
Another way to say this is as follows:  Suppose we
find our solution at tree level at some fixed value of fluxes.
We can choose our coordinates of moduli $t_i$ such that the
K\"ahler potential will have an expansion
\begin{equation}
K=t_i{\overline t_i} +a_{ij}t_i{\overline t_j}f(t,{\overline t}) \,,
\end{equation}
where $t_i={\overline t_i}=0$ is the solution.  Quantum
corrections to $\partial_i K$ evaluated at $t_i={\overline t_i}=0$
will affect the solution only by terms which are purely holomorphic,
i.e., corrections of the form
\begin{equation}
\delta K= \delta f(t) +c.c. \,.
\end{equation}
But this can be reabsorbed to the definition of the holomorhic
section of $W$, i.e. $W\rightarrow exp(-\delta f(t))W$ will get rid
of it, without affecting our solution.

\subsection{Dirac Quantization condition}

Throughout this paper, our basic strategy for finding vacua is to start
from the effective four-dimensional superpotential induced by the fluxes
and then find its critical points as described in the previous subsection.
On top of that, we impose all known string consistency conditions which
are not captured by the four-dimensional supergravity description. An
example for such a condition is the tadpole cancellation condition, which
crucially puts a bound on the total amount of flux that can be turned on.
We will discuss this condition in the next subsection.

Another condition which cannot be seen purely within supergravity is the
Dirac quantization condition for the fluxes. This constraint arises
from the requirement that the quantum mechanics for various brane probes
charged under these fluxes be consistent.\footnote{It has been argued
\cite{uranga} that other consistency conditions such as the tadpole
cancellation can also be seen from the brane probe point of view.} Flux
quantization is notoriously delicate to analyze in topologically non-trivial
backgrounds, and this is even more true in the presence of orientifolds.
One potential subtlety is related to the so-called Freed-Witten anomaly
\cite{wittenflux,frewi,dmw}, whose full consequences in orientifolds has,
to our knowledge, not been rigorously worked out even in the supergravity
regime \cite{dfm} (but see \cite{zwirner}). Conceivably, one could
translate these constraints to the worldsheet and check whether they
are satisfied in our non-geometric models.

Another subtlety of flux quantization in orientifolds was pointed
out in \cite{frpo,mike}. \footnote{It is possible that this
condition is subsumed in the complete analysis of the Freed-Witten
anomaly for orientifolds. We discuss it here as if it were
independent.} To discuss this, let us consider a manifold $X$
together with an involution $\sigma$, by which we wish to dress
world-sheet parity to construct an orientifold model. One usually
calls $X$ the ``covering space'' of the orientifold $X/\sigma$. It
can then happen that the quotient space $X/\sigma$ has cycles that
are not inherited from $X$ (see \cite{frpo} for examples). Indeed,
consider a $p$-cycle $C\subset X$ which is mapped to itself by
$\sigma$, but meets the fixed point locus of $\sigma$ in a
lower-dimensional cycle. Then $C/\sigma$ is a $p$-cycle of
$X/\sigma$ which is represented in $X$ by $C/2$.

The Frey-Polchinski puzzle arises \cite{frpo} when turning on a
$p$-form flux $F$ in the orientifold. Naively, one would require
that the periods over any $p$-cycle in $X/\sigma$ be integral. In
particular $\int_{C/\sigma} F \in\zet$. This means that in the
covering space, $\int_C F$ is an even integer. On the other hand,
orientifolding can be viewed simply as gauging world-sheet parity
dressed by $\sigma$, and this point of view only requires that $F$
be integral on $X$ and invariant under $\sigma$ (or anti-invariant,
depending on the intrinsic parity of $F$).

The conundrum was resolved in \cite{frpo} in favor of the covering space
point of view. Namely, at least for a single cycle $C$, one can project any
(even or odd) integral flux configuration. The naive lack of integrality
in the quotient space is repaired by noticing that the fixed loci of $\sigma$,
the O-planes, carry discrete versions of the $p$-form fluxes. Those
discrete fluxes contribute to $\int_{C/\sigma} F$, and make it integral
independent of the parity of $\int _C F$.

While this argument appears to work for a single cycle at a time, there are
mutual consistency conditions between different cycles. (Because the discrete
fluxes at the O-planes require an overall choice.) We suspect that this
condition must appear in the covering space as an obstruction to choosing
a gauge such that the gauge potential of $F$ is invariant under $\sigma$,
and not just the flux $F$ itself.

It would be important to understand this better, however we believe
that this subtlety does not affect our results: As far as discrete
NS-NS fluxes are concerned, we can argue from the world-sheet
perspective. Discrete NS-NS fluxes correspond to discrete choices in
the orientifold action on the NS-NS sector. But as is evident from
the analysis in section \ref{lgmodels}, our orientifolds do not admit
such discrete choices. Therefore, following \cite{frpo}, fluxes can
have either parity in the covering space, and there can be no
consistency condition which exist when such choices are possible. It
is natural to believe the same holds for R-R flux (which would also
be natural from the viewpoint of S-duality).

To summarize, we will simply impose the Dirac quantization condition on
the fluxes in the covering space of the orientifold. Namely, we require that
the integral of Ramond-Ramond and NS-NS flux through any three-cycle be integer,
\begin{equation}
\int_\Gamma H_{RR}  \in \zet \,,
\qquad\qquad
\int_\Gamma H_{NS} \in \zet \,,
\end{equation}
for any $\Gamma\in H_3(M;\zet)$. (We work
with a normalization for the fluxes in which the periods are directly
integer).
The compatibility with the orientifold is simply the invariance condition
\begin{equation}
\int_\Gamma H_{RR} = \int_{\sigma(\Gamma)} H_{RR}\,,
\qquad\qquad
\int_{\Gamma} H_{NS} = \int_{\sigma(\Gamma)} H_{NS}\,.
\end{equation}

\subsection{Tadpole cancellation condition}
\label{tadpolecanc}

Geometrically, the tadpole cancellation condition in type IIB reads
\footnote {A more rigorous description of the type IIB tadpole
cancellation condition can be obtained in terms of twisted K-theory.
Such a description is addressed in \cite {dfm}.}
\begin{equation}
\eqlabel{tadpole} \int_M H_{RR}\wedge H_{NS} + N_{D3} =
Q_3(\text{O-plane})\,,
\end{equation}
where the first term is the contribution of (integrally normalized)
R-R and NS-NS three-form fluxes, $N_{D3}$ is the number of wandering
D3-branes in the geometry, and $Q_3(\text{O-plane})$ is the D3-brane
charge of the orientifold plane(s). All charges are measured in the
covering space, in which a single D3-brane contributes one unit. So
eg, for the T-dual of type I compactification, there are $64$
O3-planes with total three-brane charge $32$ in our units.

Our goal is to derive the CFT equivalent of the tadpole cancellation
condition. Note that in our non-geometric model $M$, we cannot readily
evaluate equation \eqref{tadpole}. The reason is that while we know
explicitly the charge of the O-plane in terms of the LG basis of
B-branes $\Lambda_n$, or the overlaps with the closed string R-R
ground states, we do not know which one of these charges should we
identify with a D3-brane.

The LG analogue of \eqref{tadpole} can be obtained by looking at
models that are continuously connected with geometry. In the
geometry, we can identify the charges appearing in \eqref{tadpole}
in the large radius limit, and phrase them in CFT language. An
important property of the tadpole cancellation condition is that it
is a topological condition and hence does not depend on the moduli
we vary to reach the LG point. The tadpole cancellation condition
will therefore take the same form, no matter at what point in
the moduli space it is phrased.

For some simple cases, the comparison between Gepner model
orientifolds and geometry was successfully done in \cite{bhhw}. We
review this comparison and extend the check to orientifolds of the
quintic involving permutations in the appendix \ref{quintic2}. This
will be a useful check on the methods used in this section to obtain
the tadpole cancellation condition of our non-geometric model in CFT
language.

\subsubsection{Application to the non-geometric torus orbifold}

As mentioned in section \ref{lgmodels}, we can view the non-geometric
LG/Gepner model $1^9$ as a $\zet_3\times\zet_3$ orbifold of
$T^6=(T^2)^3$. The idea to identify the tadpole contribution due to
fluxes in the non-geometric model is to first identify this
contribution in $T^6$. This we can do throughout the moduli space
because, being topological, it is locally constant over the moduli
space and we know what it is at large volume, where it can be
expressed in supergravity. We then translate this knowledge into a
LG language. At this point, we forget that there ever was a
geometric interpretation, and simply track the flux contribution to
the tadpole as we orbifold the LG model for $T^6$ to obtain the
non-geometric model of our interest.\footnote{We point out that for this
procedure to be successful, it is crucial that the $T^6/\zet_3\times
\zet_3$ orbifold has no B-type charges from the twisted sectors.}

The flux contribution to the D3-brane tadpole on $T^6$ is simply
that if we turn on one unit of R-R flux through cycle $A$ and one
unit of NS-NS flux through cycle $B$, then the contribution to the
tadpole is precisely one unit of D3-brane charge for every
intersection point. In other words the R-R charge generated by the
fluxes is
\begin{equation}
\eqlabel{basic}
[{\rm Flux}] = (A\cap B) [pt.] \,,
\end{equation}
where $[pt.]$ is the class of a point on $T^6$, where a D3-brane can
sit. In this formula, we can give a world-sheet interpretation to
the intersection product $A\cap B$, because it can be computed as
$\tr(-1)^F$ in the open string sector between D-branes wrapped on
the corresponding cycles.

To give a world-sheet (LG) interpretation to $[pt.]$, we use the
Calabi-Yau/LG correspondence for branes, which we have reviewed in
the appendix. We start from the LG model for a single $T^2$,
\begin{equation}
(W=x_1^3+x_2^3+x_3^3)/\zet_3 \,.
\end{equation}
Under the canonical CY/LG correspondence, the branes $(\Lambda_0,\Lambda_1,
\Lambda_2)$ (see section \ref{lgmodels}) arise in large volume by
restricting to the elliptic curve the ``exceptional collection''
$\wedge^n\Omega(n)$ from the ambient $\projective^2$ (where $\Omega$
is the cotangent bundle of $\projective^2$). It is simple to compute
the large volume charges of these bundles. In terms of their Chern
classes,
\begin{equation}
{\rm ch}_i(\Lambda_n) = B_{in} =
\ff{1&-2&1 \\ 0 &1 &-1} \,.
\end{equation}
In words, $\Lambda_0$ corresponds to a D2-brane wrapping the
whole $T^2$, $\Lambda_1$ corresponds to a bound state of $2$
anti-D2-branes and 3 D0-branes, and $\Lambda_2$ to a bound state
of one D2-brane and 3 anti-D0's. Here, the factor of $3$ comes from the
fact that the (hyper-)plane class $H$ of $\projective^2$ intersects
the elliptic curve $\{x_1^3+x_2^3+x_3^3=0\}\subset\projective^2$ in
three points. The LG monodromy $\Lambda_n\to\Lambda_{n+1}$ when acting
on the large volume charges looks as
\begin{equation}
A =
\ff{-2&-3\\1&1} \,.
\end{equation}
As is by now familiar, the $\Lambda_n$ are not a minimal charge basis
and do not contain a D0-brane. Such a minimal basis can be constructed
using permutation branes. Specifically, the D0-brane, which has charges
${\rm ch}(\text{D0})=(0,1/3)^T$ in our large-volume basis, arises in the
charge orbit
\begin{equation}
(1-A)^{-1} B =
\ff{0&-1&1\\\frac 13 &\frac13 &-\frac23} \,.
\end{equation}
Going back to the LG model, we note that the LG charges of the
$\Lambda_n$ in the $1^3$ model are ($\om\equiv\ee^{2\pi\ii/3}$,
$k=1,2$)
\begin{equation}
\langle \Lambda_n|k\rangle = (1-\om^k)^3\om^{kn} \,,
\end{equation}
while the charges of the set containing the point on $T^2$ are
\begin{equation}
\eqlabel{ptch}
\langle[pt.]_{T^2}|k\rangle = (1-\om^k)^2 \,.
\end{equation}
(Again, these are represented by permutation branes.)

Let us now take three copies of this LG model for $T^2$. We get
three exceptional collections $\Lambda_n^{(j)}$, where $j=1,2,3$
labels the $T^2$ factor. The point on $T^6$ is of course the
tensor product of points on the $T^2$'s, and so has charges
\begin{equation}
\langle[pt.]_{T^6}|k^{(1)}k^{(2)}k^{(3)}\rangle=
\prod_{j=1}^3 (1-\om^{k^{(j)}})^2 \,.
\end{equation}
Here, $k^{(j)}=1,2$ label the appropriate R-R ground states in the
$T^2$'s. Now we forget the geometric interpretation and state that every
intersection point (measured by $\tr(-1)^F$) between the cycle through
which we put NS-NS flux and the cycle through which we put R-R flux
contributes in the class $\prod_{j=1}^3 (1-\om^{k^{(j)}})^2$.
We can then orbifold the $T^6$ by $\zet_3\times\zet_3$ as described in
section \ref{lgmodels}. This has the effect of projecting the $k^{(j)}$,
$\Lambda_n^{(j)}$ so that a single set remains. This can be identified
as the set $\Lambda_n$ from section \ref{Bbranes}. In the orbifold then,
the tadpole contribution will be
\begin{equation}
\langle[{\rm Flux}]|k\rangle = (A\cap B)(1-\om^k)^6 \,,
\end{equation}
which can also be written as
\begin{equation}
[{\rm Flux}] = (A\cap B) \frac{[\Lambda_1]-[\Lambda_2]}9 \,.
\end{equation}
This is to be compared with the charges of the orientifold planes
\eqref{och}. Namely, we conclude from this analysis that
the tadpole cancellation condition between O-plane and fluxes
in the non-geometric model $1^9$ is (assuming no background D-branes
are present)
\begin{equation}
\eqlabel{proposal}
\int H_{NS}\wedge H_{RR}
= \begin{cases}
12 & \text{for orientifold action $\sigma_1$\,,}\\
-4 & \text{for orientifold action $\sigma_3$\,.}
\end{cases}
\end{equation}
Here, we have taken the O-plane charges for the orientifold
action involving one or three permutations, from \eqref{och}.
The additional factor of four is the contribution from the
uncompactified space-time directions. (The general formula
is $2^{d/2}$ for a d-dimensional space-time, and evaluates
to $32$ for $d=10$, and $4$ for $d=4$.)

\section{Solutions}
\label{solutions}

We are now ready to present some explicit examples in which all moduli
are stabilized by fluxes along the lines we have sketched in the introduction.
Since the foregoing two sections have been quite detailed and technical, we
will begin by rewriting explicitly the equations that we are to solve.

\subsection{Summary of the conditions}
\label{simpleansatz}

We have seen that unbroken supersymmetry requires that given
integral three-form fluxes $H_{RR}$ and $H_{NS}$ the complex
structure of $M$ and the dilaton must be adjusted so that
\begin{equation}
\eqlabel{adjust}
G=H_{RR}-\tau H_{NS} \in H^{2,1}(M) \oplus H^{0,3}(M) \,.
\end{equation}
Fluxes which have a non-trivial component along the $(0,3)$ direction lead to AdS
spaces, while fluxes with only $(2,1)$ components gives rise to Minkowski space
solutions. Except for some brief comments, we will mostly be interested in choosing
the complex structure, and trying to find a dilaton and an integral flux which is
supersymmetric for those values of the moduli. In this interpretation, the equations
\eqref{adjust} are simply linear equations in the flux quantum numbers, and at first
they are rather easy to solve.

The problem becomes more interesting when we also
impose the tadpole cancellation condition,
\begin{equation}
\eqlabel{tadcan}
\int H_{RR}\wedge H_{NS} = \frac{1}{\tau-\bar\tau}\int G\wedge\bar G
= 12 - N_{D3} \,,
\end{equation}
where $12$ is the contribution from the orientifold plane for
the orientifold action, $\sigma_1$, on which we concentrate from now on
(we have not been able to find any solutions for, $\sigma_3$).  Here
$N_{D3}$ the number of D3-branes that we might want to allow. The LHS of equation
\eqref{tadcan} is {\it quadratic} and {\it positive definite} in the flux
quantum numbers. Moreover, the fluxes being quantized leads to a quantization of
the LHS of \eqref{tadcan}, and it is at priori not clear whether the smallest
quantum consistent with \eqref{adjust} will be sufficiently small.

As we will see, the simplest solutions of \eqref{adjust} do in fact {\it not
satisfy} \eqref{tadcan}. We will nevertheless present this simplest
ansatz first and then improve on it, eventually exhibiting a supersymmetric
flux satisfying all requirements.

\subsection{Ansatz}

Recall that we have introduced an integral basis $\{\Gamma_\n\}$ of the
lattice of A-cycles which are labeled by the first $170$ non-negative
integers in binary notation with $9$ digits, $\n=(n_1,n_2,\ldots,n_9)$,
$n_i=0,1$. We can then introduce a set of ``three-forms'' $\gamma_\n$ which
are Poincar\'e dual to the $\Gamma_\n$, \ie,
\begin{equation}
\int_{\Gamma_\m} \gamma_\n = \Gamma_\m\cap \Gamma_\n = {\bf I}_{\bf mn} \,,
\end{equation}
where ${\bf I}_{\m\n}$ is the intersection form \eqref{inters}. For convenience,
we also introduce a ``dual basis'' $\gamma^\n$ of three-forms, defined by the
condition
\begin{equation}
\int_{\Gamma_\m} \gamma^\n = \delta_\m^\n \,.
\end{equation}
Clearly, $\gamma^\m= {\bf I}^{\m\n}\gamma_\n$ where ${\bf I}^{\m\n}$ is
the inverse of ${\bf I}_{\m\n}$, and $\int \gamma^\m\wedge\gamma^\n =
{\bf I}^{\m\n}$.

Also recall the LG description of $H^3(M)$ according to which
harmonic forms are represented by R-R sector ground states which are
labeled by a set of nine integers
\begin{equation}
{\bf l}=\mid l_1,\dots,l_9\rangle \qquad {\rm with } \qquad l_i=1,2
\qquad {\rm and } \qquad \sum_{i=1}^9 l_i = 0 ~{\rm mod} ~3\,.
\end{equation}
The Hodge decomposition of $H^3(M)$ at the Fermat point is displayed
in table \ref{hodge}.
\begin{table}[t]
\begin{center}
\begin{tabular}{|c|l|l|}
\hline
$\sum l_i$   & spans\\
\hline
$18$  & $H^{0,3}$\\
$15$ &  $H^{1,2}$ \\
$12$ & $H^{2,1}$ \\
$9$ & $H^{3,0}$ \\\hline
\end{tabular}
\caption{Landau Ginzburg representation of $H^3(M)$ at the Fermat
point.} \label{hodge}
\end{center}
\end{table}
The pairing between homology and cohomology is given by
\begin{equation}
\eqlabel{ints} \int_{\Gamma_{\bf n}} \Omega_{\bf l} = B_{\bf l}
\;\om^{{\bf n}\cdot {\bf l}} \qquad {\rm with } \qquad {\bf
n}\cdot {\bf l}=\sum n_i l_i\,.
\end{equation}
Here $B_{\bf l}$ is an ${\bf l}$-dependent constant which will eventually
drop out of our equations.

There are now two ways to parameterize the general solution to \eqref{adjust}
(which, again, we view as an equation for the flux, fixing the complex
structure at the Fermat point). One is to start from the integral ansatz
\begin{equation}
\eqlabel{ansatz1}
G = H_{RR}-\tau H_{NS} = \sum N^{\bf n} \gamma_{\bf n} - \tau \sum M^\n\gamma_\n\,,
\end{equation}
and then impose
\begin{equation}
\eqlabel{ai} \int G \wedge \Omega_{\l} =0\qquad \text{for all
$\l$ with $\sum l_i = 12,18$ }
\end{equation}
as a constraint on the flux quantum numbers $N^{\bf n}$, $M^\m$, The
alternative is to start from an ansatz
\begin{equation}
\eqlabel{ansatz2}
G = \sum_{\sum l_i=12,18}A_\l\; \Omega_\l \,,
\end{equation}
and then adjust the coefficients $A_\l$ in such a way that $G$ has all
integral periods
\begin{equation}
\eqlabel{hasall}
\int_{\Gamma_\n} G = N_\n -\tau M_\n \,.
\end{equation}
The two parameterizations are clearly related, by $N^\n = {\bf I}^{\n\m}N_\n$
and $M^\m={\bf I}^{\m\n}M_\m$ (where all $N$'s and $M$'s are integer).

Even more explicitly, using \eqref{ints} the conditions \eqref{ai} reduce to
\begin{equation}
\eqlabel{obtain}
\sum_{\bf n} (N^{\bf n} - \tau M^{\bf n}) \om^{{\bf
n}\cdot {\bf l}} =0  \qquad {\rm where} \qquad \sum l_i=12,18\,.
\end{equation}
Note that anyone of these equations implies that $\tau$ is of the form
(this also follows alternatively from \eqref{hasall})
\begin{equation}
\eqlabel{aai} \tau = \frac{a \om + b}{c \om +d} \,,
\end{equation}
where $a,b,c,d$ are integers. As a result the value of $\tau$ is
constrained. As we will see below, solutions $\tau=\om$ (where
$\omega$ is a third root of unity), which
corresponds to one of the cusps in the fundamental domain of the
torus, can be explicitly constructed\footnote{Solutions in which
$\tau = i$, which would correspond to another fixed point of the
fundamental domain, are not allowed since they cannot be written in
the form \eqref{aai}.}.

Finally, we should also ensure that our flux is invariant under the
orientifold action. To implement this, we study the action of
$\sigma_i$ on our basis $\Gamma_{\bf n}$. (We should restrict from
now on to $i=1,3$, since only in that case can fluxes be supersymmetric
with respect to the orientifold plane at all.) By using $\sigma_i(
\Gamma_{\bf n})=\Gamma_{\sigma_i({\bf n})}$ and our choice of basis
explained in section \ref{Abranes}, it's not hard to find the expansion
\begin{equation}
\sigma_i(\Gamma_{\bf n}) = {S_i}_{\bf n}^{\;\bf m} \Gamma_{\bf m} \,,
\end{equation}
where $S_i$ is a $170\times 170$-dimensional matrix. The invariance
condition on the flux quantum numbers in \eqref{ansatz1} can then be
written as
\begin{equation}
\eqlabel{easy} N^\m{S_i}_{\n}^{\;\m} = N^{\m}\qquad\qquad
M^\n(S_i)_{\n}^{\;m} = M^{\m}\,.
\end{equation}
These equations are easy to solve over the integers and allow to
rewrite equation \eqref{obtain} in terms of $2h_{21}^++2=128,96$
independent flux quantum numbers for $\sigma_1$ and $\sigma_3$,
respectively.

It is similarly simple to impose invariance under the orientifold
on the ansatz \eqref{ansatz2}. In either way it turns out that the
invariance condition on the fluxes is not a severe restriction on
the spectrum of possible solutions.

\subsection{One flux component}

A simple solution to the supersymmetry constraints (which, however,
does not satisfy the tadpole cancellation condition) is provided by
a flux proportional to the $\overline\Omega$ component corresponding
to the R-R ground state with $|{\bf l}\rangle=|222222222\rangle$. In
this case
\begin{equation}
\eqlabel{previous} \int_{\Gamma_{{\bf n}}} \overline\Omega = A\;
\om^{2|{\bf n}|}\qquad {\rm where} \qquad |{\bf n}|=\sum_i n_i\bmod
3\,.
\end{equation}
Here $| \bf n | $ takes three different values, and $A$ is some
constant. Taking into account that $1+\om+\om^2=0$ it turns out that
the flux numbers are determined by four integers only, which we
denote by $N_0, N_1, M_0$ and $M_1$, where the index on the flux
numbers denotes the value of $|{\bf n}|$. These integers are
constrained to satisfy the determining equations for the dilaton and
the parameter $A$
\begin{equation}
\eqlabel{string} \tau = \frac{N_0 - \om N_1 }{M_0 - \om M_1 } \qquad
{\rm and } \qquad A = N_1 - \tau M_1\,.
\end{equation}
It is not hard to find that the contribution of this flux configuration
to the tadpole is given by
\begin{equation}
\eqlabel{words}\int H_{NS}\wedge H_{RR} = M_{\bf n} {\bf I}^{\bf nm}
N_{\bf m} = 27 \left(N_1 M_0 - N_0 M_1\right)\,.
\end{equation}
For integer values of $M_0,N_0,M_1,N_1$, this result \eqref{words}
is clearly in excess of the tadpole contribution from the O-plane
\eqref{tadcan}, but this will be improved upon shortly.

It is an instructive check to derive the result \eqref{words} in a
different basis of three-cycles, called the homogeneous basis. This
basis is spanned by the cycles dual to the R-R sector ground states
according to
\begin{equation}
\int_{C^{\bf l}} \Omega_{{\bf l}'} = \delta^{\bf l}_{\bf l'}\,.
\end{equation}
Note that under complex conjugation the set of integers
characterizing a R-R sector ground state transforms by interchanging
$l_i=1$ and $l_i=2$. As a result we define the index $\bar  l_i =
1+l_i~ {\rm mod} ~2$ and have
\begin{equation}
\int_{C^{\bf l}} \overline \Omega_{\bf \overline l'} = \delta^{\bf
l}_{{\bf l'}}\,,
\end{equation}
Indices are raised and lowered, again, with the help of the
intersection matrix
\begin{equation}
{\bf J}_{\bf l l'}= \int \Omega_{\bf l} \wedge \overline \Omega_{\bf
\overline  l'} = \alpha \delta_{\bf l, l'}\,,
\end{equation}
which turns out to be diagonal. Here $\alpha=i 27 \sqrt{3}$ is a
normalization constant. This expression is useful since it provides
an alternative derivation of the intersection matrix ${\bf I}_{\bf
mn}$ after transforming back to the basis spanning the integral
lattice using \eqref{ints}. We now apply the Riemann bilinear
identity and obtain
\begin{equation}
 \int G \wedge \overline G= {\bf J}_{\bf l l'} \int_{C^{\bf l}} G \int_{C^{\bf
\overline l'}} \overline G\,.
\end{equation}
In case that one flux component in the (0,3) direction is turned on,
{\it i.e.} if $G= A \overline \Omega$ this yields
\begin{equation}
\eqlabel{axxxvi} \int H_{NS} \wedge H_{RR}=\frac{1}{\tau - \bar \tau}
\int G \wedge \overline G = 27 \left(N_1 M_0 - N_0 M_1\right)\,,
\end{equation}
where we have used the result for $\tau$ and the quantization
condition for $A$ in equation \eqref{string}. Here, then, is an
alternative derivation of \eqref{words}. The homogeneous basis is
practical since it results in a diagonal intersection matrix.
However, flux quantization becomes cumbersome in this basis.

Note that the same line of reasoning shows that a minimal
non-trivial contribution to the tadpole of 27 is present each time a
single component in any of the $(2,1)$ directions is present. Thus
no flux along a single direction in $H^{2,1}(M) \oplus H^{0,3}(M)$
provides a solution of the tadpole cancellation condition. It remains
to show that the combination of several flux components will reduce
the minimal non-trivial value of the tadpole. We do this in the next
subsection.

\subsection{The general supersymmetric flux}

One may attempt to improve on the previous flux configuration, with
smallest tadpole contribution of $27$, by turning on some (small)
number of supersymmetric flux components in the homogeneous
ansatz \eqref{ansatz2},
\begin{equation}
\eqlabel{axxx} G  = \sum_{i=1}^N A_i \Omega_{\l^{(i)}} \,.
\end{equation}
As we have noted, we need to make the flux invariant under the
orientifold. In other words, we have only $b_3^+/2= h_{21}^++1$
independent fluxes we can turn on, where $b_3^+$ and $h_{21}^+$
are tabulated in table \ref{inv}. We denote the number of components
with $N$. As a result the flux spans an
hyperplane. We are interested in the sublattice created from the
intersection of this hyperplane with the integral lattice given by
$H^3(M,\zet)$, {\it i.e.}
\begin{equation}
\left(H^{2,1}(M)\oplus H^{0,3}(M) \right)\cap H^{3}(M,\zet)\,.
\end{equation}

Note that the contribution to the tadpole can be succinctly written
in the form ($\alpha\equiv\ii 3^{7/2}$)
\begin{equation}
\eqlabel{axxxiv} \frac{1}{\tau -\bar \tau} \int G \wedge \overline G
= \frac{\alpha}{\tau - \bar \tau}\sum _{i=1}^N \mid A_i\mid^2 \,,
\end{equation}
where the coefficients $A_i$ have to be chosen so that the flux is
integrally quantized. In order to impose integrality we note that
\eqref{ai} and \eqref{axxx} implies
\begin{equation}
\eqlabel{axxxiii} \sum_{i=1}^N A_i \Omega_{\l^{(i)}} =\sum_{\bf n}
\left( N^{\bf n} - \tau M^{\bf n} \right) \gamma_{\bf n}\,,
\end{equation}
which becomes
\begin{equation}
\eqlabel{diophantine} \int_{\Gamma_{\bf n}} G = \sum_j A_j
\omega^{m_j} = N_{\bf m} - \tau M_{\bf m}
\end{equation}
after integrating over the integral basis. Here ${\bf m}=
(m_1,\ldots) = (\n\cdot \l^{(1)},\ldots)$, \ie, $m_j=0,1,2$.

Note that by squaring the two sides \eqref{axxxiii} one obtains
\begin{equation}
\eqlabel{axxxv} \frac{\alpha }{\tau - \bar \tau } \sum_{i=1}^N \mid
A_i\mid^2 = N_{\bf n} {\bf I}^{\bf nm} M_{\bf m}\,.
\end{equation}
As a result once \eqref{diophantine} is imposed the contribution to
the tadpole given by \eqref{axxxiv} will always be integral.
However, due to \eqref{obtain} not all the flux quanta are
independent. Consequently even though the entries of the
intersection matrix ${\bf I}_{\bf mn}$ are $\pm 1$ the flux numbers
will appear on the right hand side of \eqref{axxxv} with a certain
multiplicity. This multiplicity is the origin of the factor 27 on
the right hand side of \eqref{axxxvi}.

Since the D3-brane charge originating from the orientifold plane is
12 the minimal non-trivial contribution of the three-form fluxes can
maximally be 12 in order to lead to a vanishing total tadpole.
Achieving this while at the same time satisfying the set of
Diophantine equations \eqref{diophantine} is highly non-trivial and
the existence of solutions is not a priori guaranteed. Below we show
the existence of solutions by presenting a set of explicit examples.

\subsection{Some sample solutions}

We have seen that the simplest supersymmetric flux, $G\propto \overline
\Omega$ makes a minimal contribution to the tadpole of $27$.
It is not hard to see that by turning on $2$ flux components ($N=2$ in the
notation of the previous subsection), we can reduce this value to $18$,
which however is still too large. Turning on more components makes the
equations increasingly cumbersome and it is not easy to find the general
integral solution by working with the ansatz \eqref{ansatz2}. One may
instead attempt to work with the integral ansatz \eqref{ansatz1}, although
this has the disadvantage that the tadpole contribution is far less
controlled.

In any event, after a tedious but in the end serendipitous search,
we have found solutions satisfying all requirements, including the
tadpole cancellation condition. Namely, we have found supersymmetric
flux configurations which are invariant under the orientifold action
$\sigma_1$ and make a tadpole contribution of $12$, or $8$. (We have
not found any solutions consistent with the orientifold $\sigma_3$.)

Let us write down explicitly three examples of solutions we have found, all
corresponding to an axio-dilaton combination
\begin{equation}
\tau = \omega\,,
\end{equation}
resulting in a string coupling constant, $g_s = 2/\sqrt{3}$. Let us
write out these solutions in both the integral basis and in the
homogeneous basis. Namely, the flux
\begin{equation}
\begin{split}
H_{RR}^1 =&
- \gamma_{0 0 0 0 1 0 1 0 1}
+ \gamma_{0 0 0 0 1 0 1 1 0}
+ \gamma_{0 0 0 0 1 1 0 0 1}
- \gamma_{0 0 0 0 1 1 0 1 0}
+ \gamma_{0 0 0 1 0 0 1 0 1} \\ &
- \gamma_{0 0 0 1 0 0 1 1 0}
- \gamma_{0 0 0 1 0 1 0 0 1}
+ \gamma_{0 0 0 1 0 1 0 1 0}
+ \gamma_{0 0 1 0 0 0 1 0 1}
- \gamma_{0 0 1 0 0 0 1 1 0} \\ &
- \gamma_{0 0 1 0 0 1 0 0 1}
+ \gamma_{0 0 1 0 0 1 0 1 0}
+ \gamma_{0 0 1 1 0 0 1 0 1}
- \gamma_{0 0 1 1 0 0 1 1 0}
- \gamma_{0 0 1 1 0 1 0 0 1} \\ &
+ \gamma_{0 0 1 1 0 1 0 1 0}
+ \gamma_{0 0 1 1 1 0 1 0 1}
- \gamma_{0 0 1 1 1 0 1 1 0}
- \gamma_{0 0 1 1 1 1 0 0 1}
+ \gamma_{0 0 1 1 1 1 0 1 0}
\\
H_{NS}^1 =&
+ \gamma_{0 0 0 0 0 0 1 0 1}
- \gamma_{0 0 0 0 0 0 1 1 0}
- \gamma_{0 0 0 0 0 1 0 0 1}
+ \gamma_{0 0 0 0 0 1 0 1 0}
+ \gamma_{0 0 0 0 1 0 1 0 1} \\ &
- \gamma_{0 0 0 0 1 0 1 1 0}
- \gamma_{0 0 0 0 1 1 0 0 1}
+ \gamma_{0 0 0 0 1 1 0 1 0}
+ \gamma_{0 0 0 1 1 0 1 0 1}
- \gamma_{0 0 0 1 1 0 1 1 0} \\ &
- \gamma_{0 0 0 1 1 1 0 0 1}
+ \gamma_{0 0 0 1 1 1 0 1 0}
+ \gamma_{0 0 1 0 1 0 1 0 1}
- \gamma_{0 0 1 0 1 0 1 1 0}
- \gamma_{0 0 1 0 1 1 0 0 1} \\ &
+ \gamma_{0 0 1 0 1 1 0 1 0}
- \gamma_{0 0 1 1 0 0 1 0 1}
+ \gamma_{0 0 1 1 0 0 1 1 0}
+ \gamma_{0 0 1 1 0 1 0 0 1}
- \gamma_{0 0 1 1 0 1 0 1 0}
\\
G^1=&H_{RR}^1-\tau H_{NS}^1 =\frac13 \Big( \Omega_{1 1 1 1 2 2 1
2 1} - \Omega_{1 1 1 1 2 2 1 1 2} - \Omega_{1 1 1 1 2 1 2 2 1} +
\Omega_{1 1 1 1 2 1 2 1 2}\Big)
\end{split}
\end{equation}
has tadpole
\begin{equation}
\int H_{RR}^1\wedge H_{NS}^1 = 12 \,,
\end{equation}
the configuration
\begin{equation}
\begin{split}
H_{RR}^2 =&
- \gamma_{0 1 0 0 1 0 1 0 1}
+ \gamma_{0 1 0 0 1 0 1 1 0}
+ \gamma_{0 1 0 0 1 1 0 0 1}
- \gamma_{0 1 0 0 1 1 0 1 0}
+ \gamma_{0 1 0 1 0 0 1 0 1} \\ &
- \gamma_{0 1 0 1 0 0 1 1 0}
- \gamma_{0 1 0 1 0 1 0 0 1}
+ \gamma_{0 1 0 1 0 1 0 1 0}
\\
H_{NS}^2 =&
- \gamma_{0 0 0 0 1 0 1 0 1}
+ \gamma_{0 0 0 0 1 0 1 1 0}
+ \gamma_{0 0 0 0 1 1 0 0 1}
- \gamma_{0 0 0 0 1 1 0 1 0}
+ \gamma_{0 0 0 1 0 0 1 0 1} \\ &
- \gamma_{0 0 0 1 0 0 1 1 0}
- \gamma_{0 0 0 1 0 1 0 0 1}
+ \gamma_{0 0 0 1 0 1 0 1 0}
\\
G^2 =& H_{RR}^2-\tau H_{NS}^2 =\frac{1}{3(1-\om)} \Big( - \Omega_{1 1
1 2 1 2 1 2 1} + \Omega_{1 1 1 2 1 2 1 1 2} + \Omega_{1 1 1 2 1 1 2
2 1} -\\ &  \Omega_{1 1 1 2 1 1 2 1 2}  + \Omega_{1 1 1 1 2 2 1 2 1}
- \Omega_{1 1 1 1 2 2 1 1 2} - \Omega_{1 1 1 1 2 1 2 2 1} +
\Omega_{1 1 1 1 2 1 2 1 2}\Big)
\end{split}
\end{equation}
contributes
\begin{equation}
\int H_{RR}^2\wedge H_{NS}^2 = 8\,,
\end{equation}
and finally, for
\begin{equation}
\begin{split}
H_{RR}^3=&
+ \gamma_{0 0 0 0 0 0 0 0 1}
+ \gamma_{0 1 0 0 0 0 0 0 1}
+ \gamma_{1 0 0 0 0 0 0 0 1}
+ \gamma_{0 0 1 0 0 0 0 0 1}
- \gamma_{1 1 1 0 0 0 0 0 1}  \\ &
- \gamma_{0 0 0 0 1 0 0 0 0}
- \gamma_{0 1 0 0 1 0 0 0 0}
- \gamma_{1 0 0 0 1 0 0 0 0}
- \gamma_{0 0 1 0 1 0 0 0 0}
+ \gamma_{1 1 1 0 1 0 0 0 0}
\\
H_{NS}^3=&
- \gamma_{0 1 0 0 0 0 0 0 1}
- \gamma_{1 0 0 0 0 0 0 0 1}
- \gamma_{1 1 0 0 0 0 0 0 1}
- \gamma_{0 0 1 0 0 0 0 0 1}
- \gamma_{0 1 1 0 0 0 0 0 1}  \\ &
- \gamma_{1 0 1 0 0 0 0 0 1}
+ \gamma_{0 1 0 0 1 0 0 0 0}
+ \gamma_{1 0 0 0 1 0 0 0 0}
+ \gamma_{1 1 0 0 1 0 0 0 0}
+ \gamma_{0 0 1 0 1 0 0 0 0}  \\ &
+ \gamma_{0 1 1 0 1 0 0 0 0}
+ \gamma_{1 0 1 0 1 0 0 0 0}
\\
G^3=& H_{RR}^3 - \tau H_{NS}^3 =\frac{1}{ 3 (1-\om) } \Big(
-\Omega_{1 1 1 2 2 2 1 1 1}
-\Omega_{1 1 1 2 2 1 2 1 1}
-\Omega_{1 1 1 2 2 1 1 2 1}\\&
+\Omega_{1 1 1 2 1 2 1 1 2}
+\Omega_{1 1 1 2 1 1 2 1 2}
+\Omega_{1 1 1 2 1 1 1 2 1}
-\Omega_{1 1 1 1 2 2 2 1 1}
-\Omega_{1 1 1 1 2 2 1 2 1}\\ &
-\Omega_{1 1 1 1 2 1 2 2 1}
+\Omega_{1 1 1 1 1 2 2 1 2}
+\Omega_{1 1 1 1 1 2 1 2 2}
+\Omega_{1 1 1 1 1 1 2 2 2}\Big)
\end{split}
\end{equation}
one finds
\begin{equation}
\int H_{RR}^3\wedge H_{NS}^3 = 12\,.
\end{equation}

It is interesting that the point $\tau=\omega$ that we have found for the axio-dilaton
corresponds to one of the cusps or fixed points in the fundamental domain of the torus.

As advertised before, these solutions correspond to Minkowski space, which follows
from the fact that the coefficient of $\Omega_{222222222}$ is zero in all solutions
we have found.

\section{The \texorpdfstring{$2^6$}{26} Gepner model}
\label{other}

We have seen in the previous section that it is rather hard to find
supersymmetric flux configurations satisfying the tadpole
cancellation condition in the $1^9$ model, and in fact we have been
extremely lucky to find any solutions at all! From our perspective,
the difficulty stems mainly from the fact that the O-plane
contribution to the tadpole is so small ($12$ or $4$, depending on
the choice of involution). One naturally wonders why this is so.
After all, our model is nothing but a torus orbifold, and for the
orientifold of $T^6$ in which world-sheet parity inverts all $6$
torus directions, there are $64$ O3-planes, with total charge $32$.
The reason we get something smaller in the LG description can be
traced back to the fact that the $1^3$ Gepner model actually
corresponds to a $T^2$ with B-field $B=1/2$. One can show that this
forces some of the O3-planes to be ``exotic'' in the sense that they
have positive charge where regular O3-planes have negative charge.
This reduces the contribution from the O-planes. However, this
observation also indicates that it should be easier to find
solutions in a model related to $T^6$ with $B=0$, for which tadpole
canceling flux configurations were for example discussed in
\cite{mike}. Indeed,  there is such a Gepner model, which is the
so-called $2^6$ model. This is also a torus orbifold
$T^6/\zet_4\times\zet_4$ (with zero B-field) with Hodge numbers
$h_{11}=0$, $h_{21}=90$. In this section, we will see that repeating
the analysis for this model has several payoffs.

\subsection{The model}

The so-called $2^6$ model is best understood as emerging from the
world-sheet superpotential
\begin{equation}
W=\sum_{i=1}^6 x_i^4 + z^2 \,,
\end{equation}
divided by a $\zet_4$ action
\begin{equation}
g: x_i \to \ii x_i\qquad\qquad z\to -z\,.
\end{equation}
The extra $z^2$ term in the superpotential might seem trivial
and indeed it can be integrated out. However, doing so, the
orbifold action $x_i\to \ii x_i$ has to be dressed by $(-1)^F$
and this is somewhat awkward to implement at the level of the
branes. It is generally recommended\footnote{Geometrically
it is more natural to add three quadratic fields $z_i^2$.} to
study LG models with the number of fields congruent to $\hat c$
modulo $2$.

Similarly to the $1^9$ model, the $2^6$ model is an orbifold
$T^6/\zet_4\times\zet_4$. So most of the previous discussion
carries over to the present case, however there are several
subtleties associated with the fact that the levels are now
even. For example, we have new choices in the orientifold
action. The canonical choice is
\begin{equation}
\sigma_0: (x_1,\ldots,x_6)\to \om(x_1,\ldots,x_6)
\qquad z\to \ii z \,,
\end{equation}
where now $\om=\ee^{2\pi\ii/8}$. The orientifold group is
$\zet_8$ and fits into the sequence
\begin{equation}
\zet_4\longrightarrow \zet_8\longrightarrow\zet_2 \,.
\end{equation}
The orientifold action can be dressed in various ways by
symmetries. The restrictions are that parity remain involutive
up to the orbifold group and we count orientifold actions as
equivalent when they differ by conjugation by a symmetry. For
example, we can have (suppressing $z\to\ii z$)
\begin{equation}
\eqlabel{various}
\begin{split}
\sigma_1&:
(x_1,x_2,x_3,x_4,x_5,x_6) \to \om(-x_1,x_2,x_3,x_4,x_5,x_6)\,,\\
\sigma_2&:
(x_1,x_2,x_3,x_4,x_5,x_6) \to \om(-x_1,-x_2,x_3,x_4,x_5,x_6)\,,\\
\sigma_3&:
(x_1,x_2,x_3,x_4,x_5,x_6) \to \om(-x_1,-x_2,-x_3,x_4,x_5,x_6)\,,\\
\sigma_4&:
(x_1,x_2,x_3,x_4,x_5,x_6) \to \om(-x_1,-x_2,-x_3,-x_4,x_5,x_6)\,,\\
\sigma_5&:
(x_1,x_2,x_3,x_4,x_5,x_6) \to \om(-x_1,-x_2,-x_3,-x_4,-x_5,x_6)\,,\\
\sigma_6&:
(x_1,x_2,x_3,x_4,x_5,x_6) \to \om(-x_1,-x_2,-x_3,-x_4,-x_5,-x_6)\,.\\
\end{split}
\end{equation}
Note that $\sigma_6=g^2\sigma_1$, $\sigma_5=g^2\sigma_2$, $\sigma_4=
g^2\sigma_3$, so these parities define the same orientifold.
One might also note that we did not have the similar option to dress
parity action with a non-trivial phase in the $1^9$ model, where all
levels are odd. In the present case, we can in addition to the phase
symmetries also consider permuting some of the variables, but these
parities are always equivalent by a change of variables to one of the
$\sigma_i$'s, perhaps with a change of superpotential. Eg,
the action $x_1\to x_2$ on $x_1^4+x_2^4$ is equivalent to
$(\tilde x_1,\tilde x_2)=(x_1-x_2,x_1+x_2)\to (-\tilde x_1,\tilde x_2)$,
with the superpotential $\tilde x_1^4 + 6 \tilde x_1^2\tilde x_2^2+
\tilde x_2^4$. The projection of moduli is given in the table \ref{26proj}
\begin{table}[ht]
\begin{center}
\begin{tabular}{|c|l|l|l|}
\hline
orientifold & $h_{21}^+$ & $b_3^+$\\
\hline
$\sigma_0$ & 90 & 182 \\
$\sigma_1$ & 60 & 122 \\
$\sigma_2$ & 50 & 102 \\
$\sigma_3$ & 48 & 96
\\\hline
\end{tabular}
\caption{Number of invariant complex structure deformations for
various orientifolds of the $2^6$ model.}
\label{26proj}
\end{center}
\end{table}
There is one further option in the orientifold action which was not
available for the $1^9$ model, namely the ``dressing by quantum symmetry''.
Recall that the quantum symmetry associated with the $\zet_4$ orbifold
is $\zet_4^*\cong \zet_4$ and measures the
twisted sector. Dressing parity by an element $\chi\in\zet_4^*$ means
that we multiply a state in the sector twisted by $g\in\zet_4$ by the
phase $\chi(g)$. Any such dressing is involutive, and those related
modulo $(\zet_4^*)^2$ are equivalent. In upshot, we have one non-trivial
dressing by quantum symmetry, and we will denote the corresponding
orientifolds by $\tilde \sigma_i$, $i=0,1,2,3$. The projection of
complex structure moduli is unchanged, whereas the projection of
K\"ahler parameters, if they were present, would be different. (See
\cite{bhhw} for examples of such situations.)

\subsection{A-branes}

In the $x^4$ minimal model, we have to divide the cake into 4 pieces, which
we call $V_0,V_1,V_2,V_3$, satisfying the relation $V_0+V_1+V_2+V_3=0$,
and having intersection matrix
\begin{equation}
{\rm id} -g = \begin{pmatrix} 1&0&0&-1\\-1&1&0&0\\0&-1&1&0\\0&0&-1&1
\end{pmatrix}\,.
\end{equation}
In the $z^2$ factor, there are only two straight wedges, which only
differ by orientation. Following the same strategy as before in the
$1^9$ model, we obtain basic A-branes $\Gamma_{[\n]}$ in the $2^6$ model,
where $\n=(n_1,\ldots,n_6,n_z)$, $n_i\equiv n_i\bmod 4$,
and $n_z\equiv n_z\bmod 2$. The orbifold equivalence is $\n\equiv\n+
(1,1,1,1,1,1,1)$. The intersection matrix in the orbifold is
\begin{equation}
(1-g)^{\otimes 6} \otimes (1-g_z)^{\otimes 6}
\bigl(1+g^{\otimes 6}\otimes g_z + (g^{\otimes 6}\otimes g_z)^2+
(g^{\otimes 6}\otimes g_z)^3 \bigr)\,.
\end{equation}
In practice, it is convenient to go to a truncated set by using the
relations satisfied by the $V_i$'s. In the $x^4$ models, the symmetry
generator and intersection matrix look, respectively,
\begin{equation}
A = \begin{pmatrix}
0 & 0 & -1\\1&0&-1\\0&1&-1
\end{pmatrix}
\qquad
I = \begin{pmatrix}
1&0&0\\-1&1&0\\0&-1&1
\end{pmatrix}\,.
\end{equation}
In the $z^2$ model, we have, $I_z=1$, $A_z=-1$. Thus, the truncated
intersection matrix of the $2^6$ model is:
\begin{equation}
\eqlabel{int26}
{\bf I} = I^{\otimes 6}\bigl(1-A^{\otimes 6} + (A^{\otimes 6})^2
- (A^{\otimes 6})^3\bigr)\,.
\end{equation}
\def\m{{\bf m}}
The formula in the closed string channel (cf., \eqref{closed}) is
\begin{equation}
\eqlabel{can26}
{\bf I}_{[\n],[\m]} = \frac{1}{4^5}
\sum_{\l} \prod(1-\ii^{-{l_i}}) \ii^{n_il_i-m_il_i}\,,
\end{equation}
where the sum is over $\l=(l_1,\ldots l_6)$ with $l_i=1,2,3$ with
$\sum l_i=0\bmod 4$ (there are $182$ of them). The formula \eqref{can26}
can be understood, from the mirror symmetry construction using
matrix factorizations, or from the wedge picture.

As in the $1^9$ model, it turns out that the first (in alphabetical order)
$182$ $\Gamma_{[\bf n]}$ with $n_i=0,1,2$ form an integral basis of the
charge lattice.

\subsection{B-branes}

The basic B-branes correspond to the tensor product of matrix
factorizations $x^4=x\cdot x^3$, on which the $\zet_4$ generator
is represented by
\begin{equation}
\gamma=\begin{pmatrix} 1&0\\0&\ii \end{pmatrix}\,.
\end{equation}
In the $z^2$ model, we only have the factorization $z\cdot z$
with $\zet_2$ generator represented by
\begin{equation}
\gamma_z = \begin{pmatrix} 1&0\\0&-1\end{pmatrix}\,.
\end{equation}
We now tensor together and orbifold, which means choosing a
representation of $\zet_4$,
\begin{equation}
g = \ii^n \gamma^{\otimes 6} \otimes \gamma_z\,.
\end{equation}
We call the resulting branes $\Lambda_n$, $n=0,1,2,3$.
There are two twisted Ramond ground states in our model,
from twisted sector $k=1$ and $k=3$. The brane charges are
(cf., \eqref{eg})
\begin{equation}
\langle\Lambda_n|k\rangle = \str g^k = 2 (1-\ii^k)^6\,.
\end{equation}
One way to see that there is no Ramond ground state for
$k=2$ is that $\str g^2=0$ for any brane. It is also easy to
see that $\Lambda_n$ is the anti-brane of $\Lambda_{n+2}$,
and consequentially $\Lambda_0$, $\Lambda_1$ form a (non-integral)
basis, with intersection form
\begin{equation}
{\bf J}_{nm} =
\frac14 \sum_{k=1,3} 2 (1-\ii^k)^6 \frac{1}{2(1-\ii^k)^6}
2(1-\ii^{-k})^6 \ii^{k(n-m)} =
\begin{pmatrix} 0 & 8 \\  -8&0\end{pmatrix}\,.
\end{equation}
A minimal basis is provided by the maximal permutation branes with
charges
\begin{equation}
\langle\Lambda_n^{(12)(34)(56)}| k\rangle
= 2(1-\ii^k)^3\,,
\end{equation}
which are related to the $\Lambda_0$, $\Lambda_1$ basis by
\begin{equation}
\eqlabel{minimal}
[\Lambda_0^{(12)(34)(56)}] = \frac{-[\Lambda_0]+[\Lambda_1]}4
\qquad\qquad
[\Lambda_1^{(12)(34)(56)}] = \frac{-[\Lambda_0]-[\Lambda_1]}4\,.
\end{equation}
By following a similar analysis as for the $1^9$ model, one
can show that these charges \eqref{minimal} are the charges
corresponding to the ``point class'' in the non-geometric
$2^6$ model, and hence are the normalization for the tadpole
contribution from the fluxes.

\subsection{O-plane charges}

To get the charges of the O-planes associated with the various orientifold
actions \eqref{various} as well as their quantum symmetry twists, we turn
again to eq.\ (5.37) in \cite{howa}. There are now two distinctions from
the $1^9$ model. First of all, we notice that for any given orbifold
element $g^k$ ($k=1,3$) with a ground state in the corresponding twisted
sector, there are {\it two} parities which square to it: If $\sigma$ is
one, then $g^2\sigma$ is the other. Thus, (5.37) becomes a sum of two
terms.

To understand the possible dressing by quantum symmetry, we have to resolve
the definition of the phase $c(\sigma)$ from eqs.\ (4.13), (4.14). Without
quantum symmetry dressing, $\chi\equiv 1$, the phase is just an overall
choice of sign of the O-plane. The non-trivial $\chi$ is defined
by $\chi(g)=\ii$, and we get the values
\begin{equation}
c(g\sigma) = -\ii c(\sigma) \qquad
c(g^2\sigma) = - c(\sigma)  \qquad
c(g^3\sigma) = \ii c(\sigma)\,,
\end{equation}
where $c(\sigma) = \pm \om^{-1}$, and $\om=\ee^{2\pi\ii/8}$. (The sign is
an overall choice, and we'll omit it.)

Now let us compute the O-plane charge associated with the canonical
orientifold action $\sigma_0$. As we said, for each orbifold element
there are two parities which square to it. Eg, for $k=1$, the eigenvalues
of $\sigma_0$ are $(\om,\om,\om,\om,\om,\om,\ii)$, while those of
$g^2\sigma_0$ are $(-\om,-\om,-\om,-\om,-\om,-\om,\ii)$.
So we obtain
\begin{equation}
\begin{split}
\langle C|1\rangle &= (1+\ii)\bigl((1+\om)^6 + (1-\om)^6\bigr) = -56 \\
\langle C|3\rangle &= (1-\ii)\bigl((1+\om^{-1})^6 + (1-\om^{-1})^6\bigr)
= -56\,.
\end{split}
\end{equation}
With dressing by quantum symmetry, we get
\begin{equation}
\begin{split}
\langle\tilde C|1\rangle &= \ii\om^{-1} (1+\ii)
\bigl((1+\om)^6 -(1-\om)^6\bigr)= -40-40\ii \\
\langle\tilde C|3\rangle &= (-\ii)(\ii\om^{-1})(1-\ii)
\bigl((1+\om^{-1})^6 - (1-\om^{-1})^6\bigr) = -40+40\ii\,.
\end{split}
\end{equation}
Continuing in this fashion, and expressing the charges in the $\Lambda_0$,
$\Lambda_1$ basis, we find the following analog of \eqref{och}
\begin{align}
\eqlabel{analog}
[O_0] &= \frac72[\Lambda_1] &[\tilde O_0] & = \frac 52(-[\Lambda_0]+[\Lambda_1]) \\
[O_1] &= \frac 32[\Lambda_0] &[\tilde O_1] & =[\Lambda_0]+[\Lambda_1] \\
[O_2] &= -\frac{[\Lambda_1]}{2}  &[\tilde O_2] & = \frac{[\Lambda_0]-[\Lambda_1]}2\\
[O_3] &= -\frac{[\Lambda_0]}{2} &[\tilde O_3] & = 0\,.
\end{align}
As before, these results should be multiplied by $4$ to get the
actual charge of the orientifold planes in space-time.

\subsection{Simple ansatz}

As we have done in the $1^9$ model, it is a useful starting point to first
study the tadpole contribution of a flux with only a $(0,3)$ component
turned on. \ie, $G=H_{NS}-\tau H_{RR} \propto \overline\Omega$.
More precisely, we set $G=A\Omega_{\l_0}$ where $\l_0=(3,3,3,3,3,3)$
in the normalization in which the intersection form is given by
\eqref{can26}, namely
\begin{equation}
\int \Omega_{\l}\wedge\overline\Omega_{\l}
= 4^5 \prod (1-\ii^{l_i}) = -2^{13} \ii\,.
\end{equation}
Imposing integrality on the fluxes means
\begin{equation}
\int_{\Gamma_{\n}} G = A \prod (1-\ii^{l_i})\ii^{n_il_i}
=N_{\n}-\tau M_{\n}\,.
\end{equation}
As in the $1^9$ model, we can parameterize this solution in terms of
just $4$ integers, $N_0$, $N_1$, $M_0$, $M_1$. We find
\begin{equation}
\tau = \frac{N_0-\ii N_1}{M_0-\ii M_1}\qquad\qquad
A = \frac{N_1-\tau M_1}{(1+\ii)^6}\,,
\end{equation}
and as a result
\begin{equation}
\eqlabel{asresult}
\begin{split}
\frac{1}{\tau-\bar\tau} \int G\wedge\bar G & =
\frac{|M_0-\ii M_1|^2}{2\ii(N_0M_1-M_0N_1)}
\;\frac{\ii 2^{13}}{|(1+\ii)^6|^2}\;
\frac{(N_1M_0-M_1N_0)^2}{|M_0-\ii M_1|^2}\\
& = 64(N_0M_1-M_0N_1)\,.
\end{split}
\end{equation}

\subsection{Tadpole cancellation}

When combining together \eqref{minimal}, \eqref{analog}, and
\eqref{asresult}, we see that we ought to use the orientifolds which
include twist by quantum symmetry in order to get O-plane charges in
the direction of a ``D3-brane'', which we have identified with
\eqref{minimal}. Moreover, we should remember to multiply the
results of \eqref{analog} by 4 to take into account the space-time
contribution.

In this way, we obtain the following tadpole cancellation condition
\begin{equation}
\eqlabel{cannot}
\int H_{RR}\wedge H_{NS} = 40, 16, 8, 0\,,
\end{equation}
respectively, for the four possible orientifolds. This equation
\eqref{cannot} cannot be satisfied by the simple ansatz and
result \eqref{asresult} used above. But there are more
complicated flux configurations which do the job.

\subsection{Sample solutions}

We have made a search for supersymmetric flux configurations in the $2^6$
model whose tadpole contribution is within the bound imposed by the
charge of at least one of the O-planes \eqref{cannot}. As anticipated,
the spectrum of possibilities is wider than in the $1^9$ model, due
to the fact that the O-plane contribution is larger. Nevertheless, most
of the solutions are still quite complicated.

As an example, the configuration
\begin{equation}
\begin{split}
H_{RR}^1=&
\gamma_{0 0 0 0 0 2}
+ 2 \gamma_{0 0 0 0 1 1}
+ 2 \gamma_{0 0 0 0 1 2}
+ \gamma_{0 0 0 0 2 0}
+ 2 \gamma_{0 0 0 0 2 1}
- \gamma_{0 0 0 1 1 0}
- \gamma_{0 0 0 1 1 2}
- \gamma_{0 0 0 1 2 0} \\ &
- \gamma_{0 0 0 1 2 2}
- \gamma_{0 0 0 2 0 0}
- \gamma_{0 0 0 2 1 0}
+ \gamma_{0 0 0 2 1 1}
+ \gamma_{0 0 0 2 2 1}
- \gamma_{0 0 1 0 0 1}
- \gamma_{0 0 1 0 0 2}
- \gamma_{0 0 1 0 2 1} \\ &
- \gamma_{0 0 1 0 2 2}
- \gamma_{0 0 2 0 0 0}
- \gamma_{0 0 2 0 0 1}
+ \gamma_{0 0 2 0 1 1}
+ \gamma_{0 0 2 0 1 2}
\\
H_{NS}^1=&
\gamma_{0 0 0 0 0 1}
+ \gamma_{0 0 0 0 1 0}
+ 2 \gamma_{0 0 0 0 1 1}
- \gamma_{0 0 0 0 1 2}
- \gamma_{0 0 0 0 2 1}
- 2 \gamma_{0 0 0 0 2 2}
- \gamma_{0 0 0 1 0 0}
- \gamma_{0 0 0 1 0 2}  \\ &
- \gamma_{0 0 0 1 1 0}
- \gamma_{0 0 0 1 1 2}
+  \gamma_{0 0 0 2 0 1}
+ \gamma_{0 0 0 2 1 0}
+ \gamma_{0 0 0 2 1 1}
+ \gamma_{0 0 0 2 2 0}
- \gamma_{0 0 1 0 0 0}
- \gamma_{0 0 1 0 0 1} \\ &
- \gamma_{0 0 1 0 2 0}
- \gamma_{0 0 1 0 2 1}
+ \gamma_{0 0 2 0 0 1}
+ \gamma_{0 0 2 0 0 2}
+ \gamma_{0 0 2 0 1 0}
+ \gamma_{0 0 2 0 1 1}
\\
G^1=&H_{RR}^1-\tau H_{NS}^1\propto
  \Omega_{1 2 1 3 2 1}
- \Omega_{1 2 2 3 1 1}
+ \Omega_{1 2 3 1 1 2}
- \Omega_{1 2 3 2 1 1}
+ \Omega_{1 3 1 2 2 1}
+ \Omega_{1 3 2 1 1 2} \\ &
- 2 \Omega_{1 3 2 2 1 1}
+ \Omega_{2 1 1 3 2 1}
- \Omega_{2 1 2 3 1 1}
+ \Omega_{2 1 3 1 1 2}
- \Omega_{2 1 3 2 1 1}
+ \Omega_{2 2 1 2 2 1}
+ \Omega_{2 2 2 1 1 2} \\ &
- 2 \Omega_{2 2 2 2 1 1}
+ \Omega_{2 3 1 1 1 2}
+ \Omega_{2 3 1 1 2 1}
- \Omega_{2 3 1 2 1 1}
- \Omega_{2 3 2 1 1 1}
+ \Omega_{3 1 1 2 2 1}
+ \Omega_{3 1 2 1 1 2} \\ &
- 2 \Omega_{3 1 2 2 1 1}
+ \Omega_{3 2 1 1 1 2}
+ \Omega_{3 2 1 1 2 1}
- \Omega_{3 2 1 2 1 1}
- \Omega_{3 2 2 1 1 1}
\end{split}
\end{equation}
has a tadpole contribution of
\begin{equation}
\int H_{RR}^1\wedge H_{NS}^1 = 40\,,
\end{equation}
exactly saturating the tadpole from $[\tilde O_0]$. Since $G^1$
does not have a $\Omega_{333333}$ component turned on, it corresponds
to a Minkowski space solution.

As another example, let us look at
\begin{equation}
\begin{split}
H_{RR}^2=&
\gamma_{0 0 0 1 0 1} + \gamma_{0 0 0 1 0 2}
+ \gamma_{0 0 0 1 1 0} + 2 \gamma_{0 0 0 1 1 1}
+ \gamma_{0 0 0 1 1 2} + \gamma_{0 0 0 1 2 0}
+ \gamma_{0 0 0 1 2 1} - \gamma_{0 0 1 0 0 1} \\ &
- \gamma_{0 0 1 0 0 2} - \gamma_{0 0 1 0 1 0}
- 2 \gamma_{0 0 1 0 1 1} - \gamma_{0 0 1 0 1 2}
- \gamma_{0 0 1 0 2 0} - \gamma_{0 0 1 0 2 1}\,,
\\
H_{NS}^2=&
\gamma_{0 0 0 1 0 0} + \gamma_{0 0 0 1 0 1} +
  \gamma_{0 0 0 1 1 0} - \gamma_{0 0 0 1 1 2} -
  \gamma_{0 0 0 1 2 1} - \gamma_{0 0 0 1 2 2} -
  \gamma_{0 0 1 0 0 0} - \gamma_{0 0 1 0 0 1} - \\ &
  \gamma_{0 0 1 0 1 0} + \gamma_{0 0 1 0 1 2} +
  \gamma_{0 0 1 0 2 1} + \gamma_{0 0 1 0 2 2}\,,
\\
G^2=& H_{NS}^2-\tau H_{RR}^2\propto
(-1 + \ii) \Omega_{3 2 2 1 1 1}
+ (1 - \ii) \Omega_{3 2 1 2 1 1}
+ 2 \ii \Omega_{3 1 3 1 1 1}
- 2 \ii \Omega_{3 1 1 3 1 1}  \\ &
- (1 - \ii) \Omega_{2 3 2 1 1 1}
 + (1 - \ii) \Omega_{2 3 1 2 1 1}
+ 2 \ii \Omega_{2 2 3 1 1 1}
- 2 \ii \Omega_{2 2 1 3 1 1}
+ (1 + \ii) \Omega_{2 1 3 2 1 1} \\ &
-(1 + \ii) \Omega_{2 1 2 3 1 1}
+ 2 \ii \Omega_{1 3 3 1 1 1}
- 2 \ii \Omega_{1 3 1 3 1 1} +
(1 + \ii) \Omega_{1 2 3 2 1 1}
- (1 + \ii) \Omega_{1 2 2 3 1 1}\,.
\end{split}
\end{equation}
This configuration has a smaller tadpole,
\begin{equation}
\int H_{RR}^2\wedge H_{NS}^2 = 16\,.
\end{equation}
As a result, if we use it in conjunction with the orientifold
$[\tilde O_0]$, the flux will not completely cancel the charge of
the O-plane. This gives the freedom to include additional D-branes
into the background. It is an interesting open question to determine
whether there exist D-branes with the correct charge but without
continuous moduli solving the tadpole cancelation condition.
However, since the solutions presented herein are at $g_s = O(1)$
the description of the properties of these D3 branes would be
difficult to control.

Finally, we have searched for solutions which have a $(0,3)$
component turned on. There are several possibilities for such
solutions giving rise to 4-d AdS space, one of which is of the form
\begin{equation}
G^3\propto
4 \ii \Omega_{3, 3, 3, 3, 3, 3}
- \Omega_{2, 2, 2, 2, 1, 1}\,,
\end{equation}
with
\begin{equation}
\int H_{RR}^3\wedge H_{NS}^3 = 40\,.
\end{equation}
(This is the only solution we do not write out in the integral basis,
as it would take several pages.)

\section{Discussion and Conclusions}
\label{conclusions}

In this paper, we have studied moduli stabilization by fluxes in LG
compactifications of type IIB string theory. We have given both a
world-sheet and a 4D effective description of fluxes in these
theories. The particular models considered are non-geometric (as
they do not have any K\"ahler moduli $h_{11}=0$) and can be
represented by orientifolds of LG models. It has been shown that the
complex structure moduli can be stabilized in terms of fluxes only,
while the tadpole cancellation condition is satisfied due to the
presence of the orientifold charge. The value of the string coupling
constant for our solutions is of the order of unity, so that our
solutions are at strong coupling and describe points in moduli space
of enhanced symmetry \cite{DeWolfe:2004ns}. This type of vacua are
of interest for model building. So for example, low energy theories
like the MSSM have a discrete R-symmetry that helps to explain the
stability of the proton.

Since our solutions are at strong coupling, our analysis heavily
relies on supersymmetry and non-renormalization theorems. The
particular vacua that have been found have ${\cal N}=1$
supersymmetry, so that only a non-vanishing $H^{0,3}$ and $H^{2,1}$
component of the flux or a linear combination thereof is allowed. It
has been shown that the classical superpotential of \cite{Gukov:1999ya}
is exact, so that our solutions persist even non-perturbatively.

Among our main results is a set of examples of totally explicit flux
configurations which are supersymmetric, invariant under the orientifold,
and satisfy the tadpole cancellation condition. Technically, these
fluxes are solutions to a large number ($\sim 100$) of linear Diophantine
equations, and a single positive definite quadratic inequality.
This type of problem was used in \cite{Denef:2006ad} to argue that
the landscape of string vacua might be so complex from the computational
complexity point of view as to preclude finding and studying individual
vacua explicitly. From this point of view, it can appear surprising that
we have found a solution in such a high-dimensional case. In fact,
in all studies of flux stabilization so far (outside of the statistical
approach), the number of moduli has been of order 1. Nevertheless, our
findings need not be viewed as being at odds with the arguments of
ref.\ \cite{Denef:2006ad}, which rely on statements about the ``generic''
problem in this class. Moreover, our problem clearly has additional
symmetry properties such as all periods being cube roots of unity.
Although we have not crucially used these structures to find the
solution, it is likely that one could.

Our models provide the first explicit examples of flux
compactifications with all moduli stabilized by fluxes only and
which have an external Minkowski space-time. The reason for this is
the absence of K\"ahler moduli. All models constructed in the
literature before, lead to AdS space-times, which is the generic
case in all geometric models.

One can ask whether the existence of 4D solutions of string theory
with all moduli stabilized and exactly vanishing cosmological
constant should have been expected. In particular, is this consistent
with the concept of ``landscape naturalness''?\footnote{We thank
S.\ Hellerman for raising the question and S.\ Giddings for a
discussion on this issue.} A possible resolution of the cosmological
constant problem is via a dense but discrete distribution of stable
and meta-stable vacua in the landscape. If zero is not a special
value, is it ``natural'' to find it on the list of allowed values?
Clearly the solution we have found adds a new angle on this question
and it would be interesting to study in more detail the
distribution of solutions of the type discussed in this paper.
In the context of supersymmetric vacua, vanishing superpotential
leads to an unbroken R-symmetry, which might make such vacua look more
natural. Some work on vacua with unbroken R-symmetry has been done in
\cite{DeWolfe:2004ns} and \cite{dine}.

One possible extension of our work would be to deform the LG model
away from the Fermat point in the complex structure moduli space.
There is one particularly interesting limit in the moduli space,
namely the mirror of the large radius limit of the corresponding
rigid Calabi-Yau manifolds. Indeed, in this limit, our models should
be related by mirror symmetry to certain type IIA vacua studied in
\cite{DeWolfe}, which found infinite families of AdS solutions with
all geometrical moduli stabilized. It would be interesting to
recover and generalize these solutions in the type IIB setup.

It would also be interesting to gain a better understanding of the
microscopic description of fluxes in non-geometric LG models. If
this can be achieved, one could also address a world-sheet
derivation of the tadpole cancellation condition and ultimately the
derivation of a dual CFT theory description of the KKLT-like AdS
vacua appearing in the string theory landscape.

\begin{acknowledgments}
We would like to thank Aaron Bergman, Nathan Berkovits, Jacques
Distler, Mike Douglas, Dan Freed, Simeon Hellerman,
Shamit Kachru, Joe Polchinski, Jessie Shelton and Wati Taylor for
valuable discussions and communications. We are all grateful to the
organizers of the 2006 Simons Workshop in Mathematics and Physics,
where much of this work was done. J.W.\ would also like to thank the
KITP for hospitality during the program on string phenomenology.

The work of K.B.\ was supported by NSF grants PHY-0612842 and
PHY-0555575. The work of M.B.\ was supported by NSF grants
PHY-0505757 and PHY-0555575. The work of C.V.\ was supported in part
by NSF grants PHY-0244821 and DMS-0244464. The work of J.W.\ was
supported in part by the Roger Dashen Membership at the Institute
for Advanced Study and by the NSF under grant number PHY-0503584.
This research was supported in part by the National Science
Foundation under Grant No. PHY99-07949.
\end{acknowledgments}

\appendix

\section{Analytic continuation}

The purpose of this appendix is to provide some background checks on
the connection between LG orientifolds and the large volume
Calabi-Yau manifolds, when it exists. In particular, we wish to
review the canonical identification of the D0-brane in the LG/Gepner
model. We also provide a non-trivial check of the formulas of
\cite{howa}, which we have used to compute the O-plane charges in
the non-geometric LG. Namely, we verify that the O-plane charges of
the exchange orientifolds of the quintic agree between the LG and CY
geometric description. This has not so far been available in the
literature, and could be useful for other purposes as well.

\subsection{Geometric Interpretation of Cardy states in Gepner models}

A general Gepner model connected with a hypersurface in weighted projective
space has an LG description with five factors\footnote{One of the $h_i$'s
could be equal to $2$, which one wouldn't see in the Gepner model.}
\begin{equation}
W = \sum_{i=1}^5 x_i^{h_i}\,,
\end{equation}
with $\sum 1-2/h_i = 3$, modded out by a $\zet_H$ symmetry, where
$H\equiv {\rm l.c.m.}(h_i)$. The corresponding hypersurface is
$X=\{W=0\}\subset \projective_{w_1,\ldots,w_5}^4$, where
$w_i=H/h_i$.

The LG description yields $H$ basic B-branes in these models, which
we'll call $\Lambda_n$, $n=0,1,\ldots H-1$. The corresponding matrix
factorizations are based on factorizing $x_i^{h_i}=x_i\cdot x_i^{h_i-1}$.
After choosing a path in K\"ahler moduli space which connects the LG model
with the large volume, we can ask for a geometric interpretation of the
$\Lambda_n$'s in terms of bundles on the corresponding hypersurface. This
was studied in great detail following the work \cite{bdlr}, and understood
in generality in \cite{dido,mayr}, using results of \cite{hiv}.
See also the recent work \cite{hhp}.
Namely, {\it following a particular path in K\"ahler moduli space},
the $\Lambda_n$ reduce to the restriction to the hypersurface of
a so-called ``exceptional collection'' of bundles on the ambient
$\projective_{w_1,\ldots,w_5}^4$. Exceptional collections are
particularly nice bases of branes to work with, and have appeared
previously \eg, in the description of mirror symmetry for Fano
varieties \cite{hiv}.

\subsection{The quintic}
\label{quintic1}

In the following we would like to consider the example of the
quintic in $\projective^4$.  In order to change the basis from LG to
large volume (LV) it is enough to determine how the charges of the
branes transform. The LG charges of these branes are \footnote{In
this section, of course, $\om=\ee^{2\pi\ii/5}$.}
\begin{equation}
\eqlabel{LG} \langle \Lambda_n | k\rangle = (1-\om^k)^5 \om^{kn}\,,
\end{equation}
where $n=0,\ldots 4$ and $k=1,\ldots 4$.

At large volume BPS charges arising from D-branes wrapping cycles in
the Calabi-Yau are determined in terms of the topology of the
embedded cycle and a choice of bundle $E$ to be
\cite{Minasian:1997mm} \cite{frewi} \cite{Green:1996dd}
\begin{equation}
Q = {\rm ch} (E) \sqrt{\hat A(T_X)}\,.
\end{equation}
Wrapping a $p$-brane on a cycle induces a D$p$-brane charge given by
the rank of $E$, for example, while lower brane charges resulting
from the expansion of the Chern character are also induced. The
Chern characters of the bundles in the exceptional collection
$\Lambda_n=\Omega^n(n)$ (here $n=0,\ldots,4$ and $\Omega$ is the
cotangent bundle of $\projective^4$) corresponding to the fractional
branes at the LG point are
\begin{equation}
\eqlabel{chern}
\begin{split}
{\rm ch}(\Lambda_0) &= 1 \\
{\rm ch}(\Lambda_1) &= -4 + H +\frac{H^2}2 +\frac{H^3}6 \\
{\rm ch}(\Lambda_2) &= 6 -3H-\frac{H^2}2+\frac{H^3}2 \\
{\rm ch}(\Lambda_3) &= -4 + 3H - \frac{H^2}2 -\frac{H^3}2  \\
{\rm ch}(\Lambda_4) &= 1 - H +\frac{H^2}2-\frac{H^3}6\,,
\end{split}
\end{equation}
where $H$ is the hyperplane class of $\projective^4$.
In matrix notation
\begin{equation}
\eqlabel{LV} B_{in}={\rm ch_i}(\Lambda_n) =
\ff{1 &-4&6&-4&1\\
0&1& -3 & 3 & -1 \\
0&\frac 12 &-\frac 12&-\frac 12&\frac 12 \\
0 & \frac 16 & \frac 12 &-\frac12&-\frac 16}\,.
\end{equation}
Combining \eqref{LV} with \eqref{LG}, we can work out the change of
basis between the LG and the LV limit.

One can then derive the matrix $A$ representing the LV counterpart
of the LG monodromy $g$ (which sends $\Lambda_n\to \Lambda_{n+1}$),
namely
\begin{equation}{\rm ch}_i(\Lambda_{n+1})= A_i^j {\rm
ch}_j(\Lambda_n) \end{equation}
with
\begin{equation}
A =
\ff{-4&-\frac{20}3 &-5 &-5 \\ 1&1&0&0\\
\frac12 & 1 & 1 & 0\\ \frac 16&\frac12 & 1 & 1}\,.
\end{equation}

The intersection matrix of the $\Lambda_n$'s is
\begin{equation}
\frac 15 \sum_{k=1}^4 (1-\om^k)^5\om^{km}
\frac{1}{(1-\om^k)^5}
(1-\om^{-k})^5\om^{-kn}
=
\ff{0& 5&-10&10&5\\
-5&0&5&-10&10\\
10&-5&0&5&-10\\
-10&10&-5&0&5\\
-5&-10&10&-5&0}\,,
\end{equation}
and again, the $\Lambda_n$'s are not a minimal integral basis of
the charge lattice. We can improve on this, as first pointed out
in \cite{dia}, by using permutation branes.

The permutation branes based on the exchange of $x_1$ and $x_2$
have LG charges
\begin{equation}
\langle  \Lambda^{(12)}_n | k \rangle = (1-\om^k)^4 \om^{kn}
= (1-\om^k)^{-1} \langle\Lambda_n|k\rangle\,,
\end{equation}
which in LV translates to the Chern characters
\begin{equation}
B_{in}^{(12)}= {\rm ch}_i(\Lambda^{(12)}_n) = \ff{0 & -1 & 3 & -3 & 1\\
0 & 0 & -1 & 2 & -1 \\
0 & 0 &-\frac12&0&\frac 12\\
\frac 15 &\frac 15 &\frac 1{30} &-\frac 7{15} &\frac{1}{30}}\,,
\end{equation}
where
\begin{equation}
B^{(12)}_{in}= \left[(1-A)^{-1} B\right]_{in}\,.
\end{equation}
The first of those
\begin{equation}
{\rm ch}(\Lambda^{(12)}_0) = {\rm ch}_3= \frac {H^3}5\,,
\end{equation}
describes a point on the quintic. Remember that $H$ is the
hyperplane class of $\projective^4$ and the quintic is in the class
$5H$, so $\Lambda^{(12)}_0$ intersects the quintic exactly once.
Note that even though this set of branes contains a D0-brane, the
$\Lambda^{(12)}_n$ are still not a minimal basis of the charge
lattice.

Continuing, the permutation branes $\Lambda_n^{(12)(34)}$, which are based
on the exchange of $x_1$ and $x_2$, and $x_3$ and $x_4$, have LG charges
\begin{equation}
\langle  \Lambda^{(12)(34)}_n | k \rangle = (1-\om^k)^3 \om^{kn}
= (1-\om^k)^{-2} \langle\Lambda_n|k\rangle
\end{equation}
and Chern characters
\begin{equation}
{\rm ch}_i(\Lambda^{(12)(34)}_n) = \left[(1-A)^{-2} B\right]_{in} =
\ff{0&0&1&-2&1\\
0&0&0&1&-1\\
-\frac 15&-\frac15&-\frac15&\frac3{10}&\frac{3}{10}\\
\frac15&0&-\frac15&-\frac7{30}&\frac{7}{30}}\,.
\end{equation}
These are now indeed a minimal basis of the charge lattice (but
do not contain a D0-brane). Their intersection form is
\begin{equation}
\frac 15 \sum_{k=1}^4 (1-\om^k)^3\om^{km}
\frac{1}{(1-\om^k)^5}
(1-\om^{-k})^3\om^{-kn}
=
\ff{0&0&-1&1&0\\
0&0&0&-1&1\\
1&0&0&0&-1\\
-1&1&0&0&0\\
0&-1&1&0&0}\,.
\end{equation}

\subsection{Identification of the D0-brane}

It follows from the previous discussion that the set of fractional branes
$\Lambda^{(12)}_n$ containing the D0-brane and the set of
Cardy-Recknagel-Schomerus branes $\Lambda_n$ containing the D6-brane are
related by the formula
\begin{equation}
\eqlabel{general}
\Lambda_n = (1-g)_{nm} \Lambda^{(12)}_m\,.
\end{equation}
Such an identification of the D0-brane in the LG model appears in fact to be
canonical and holds in particular for all hypersurfaces whose analytical
continuation has been studied so far, see, \eg, \cite{mayr,esche}.

More properly, the statement that ``one of the $\Lambda^{(12)}_n$ is
a D0-brane'' of course depends on the analytical continuation that
one has chosen to connect LG to LV. For example, encircling the LG
point leads to a cyclic permutation $n\to n+1$. One of the
consequences of this ambiguity in the context of flux
compactifications is that the statement ``the fluxes contribute to
the D3-brane tadpole'' is not invariant under all small volume
monodromies: If the D0-brane on the Calabi-Yau returns under
monodromy as a general combination of even-dimensional cycles, this
can only be consistent with tadpole cancellation if after the
monodromy, the fluxes become non-geometric and contribute in other
classes as well.

However, now we have to take into account that we are performing an
orientifold projection. In type IIB, this selects a real subspace of
the K\"ahler moduli space, which therefore eliminates some of the
possible monodromies. Moreover, as we will see in the next section
for the particular example of the quintic the orientifold projection
fixes the ambiguity completely.

In the above discussion we have seen an interesting interplay
between orientifolds, monodromies and tadpole contributions
generated by fluxes. In the present context we have used this
interplay to identify the class of a point in the LG regime but we
expect it to have implications beyond the present discussion and to
play a pivotal role in the search for the LG theories incorporating
NS-NS and R-R fluxes.

\subsection{Orientifolds of the quintic}
\label{quintic2}

We consider first the trivial involution
\begin{equation}
\sigma_0 : (x_1,x_2,x_3,x_4,x_5)\to (-x_1,-x_2,-x_3,-x_4,-x_5)\,,
\end{equation}
where, the full orientifold group consists of $g^k$ and
$g^k \sigma_0$ for $k=0\ldots 4$. To compute $\langle C|k\rangle$,
following \cite{howa}, we have to look for those elements of the
orientifold group which square to the element $g^k$ of the
orbifold group, and then compute its eigenvalues. Eg, for $k=1$,
$(g^3\sigma_0)^2=g$, with eigenvalues $(-\om^3,\ldots,-\om^3)$
and so on:
\begin{equation}
\langle C|k\rangle = (1-\om^{3k})^5\,.
\end{equation}
and using $(1-\om)^{-1}=\frac{1}{5}(4+3\om+2\om^2+\om^3)$,
we find for the class of the O-plane
\begin{equation}
\eqlabel{naive}
[O] = 3[\Lambda_0] -5 [\Lambda_2] - 5[\Lambda_3]\,,
\end{equation}
which corresponds to large volume charges $-7+5H^2$. After we recall
that the formulas in \cite{howa} are missing a factor of $4$ from
the extended directions, we see that this would correspond to a
rank $28$ bundle, which cannot be correct for tadpole cancellation for
a type I compactification on the quintic, which we would have naively
expected corresponds to this orientifold (and would hence require
a rank $32$ bundle).

A solution to this was noted in \cite{bhhw}. Recall that the
correspondence between $\Lambda_n$'s and bundles is in fact ambiguous
by the path we choose to get to large volume. In the quintic case,
the path is fixed by the orientifold projection, except at the
orbifold point. In fact,
\begin{equation}
g^2 [O] = 3[\Lambda_2]-5[\Lambda_4]-5[\Lambda_0]
\end{equation}
corresponds to large volume charges $8-4H-4H^2+\frac{7}{3}H^3$, and
gives rank 32 after we multiply by our factor of $4$. Thus, if we
modify our path by 2 LG monodromies (which is compatible with the
orientifold projection on the moduli space), we get agreement with
large volume data.

To understand that this is in fact the path we must take, we recall
that a global coordinate on the K\"ahler moduli space is the complex
structure parameter $\psi$ of the mirror quintic
\begin{equation}
y_1^5+y_2^5+y_3^5+y_4^5+y_5^5 - 5\psi y_1y_2y_3y_4y_5\,.
\end{equation}
More precisely, $\psi$ gives a five-fold cover of the moduli space,
which is usually parametrized by $z=(5\psi)^{-5}$. The LG monodromy
corresponds to $\psi\to\ee^{2\pi\ii/5} \psi$. Now, the orientifold
acts on the mirror quintic simply by complex conjugation $y_i\to
\bar y_i$, and hence restricts $\psi$ to be real.  This is a
stronger condition than requiring $z$ to be real. Navigating from
positive real $\psi$ to negative real $\psi$ in fact requires encircling
the origin of the moduli space $z=0$ twice in the positive direction (or
thrice in the negative direction).

That $g^2[O]$ does not seem to correspond to a real bundle in this
case (odd Chern classes are non-zero) is explained by the fact that
we actually end up with a non-zero NS-NS B-field (more precisely
$B=H/2$) under this analytical continuation. Namely
\begin{equation}
4 \ee^{H/2}[O] = 32 - 20 H^2
\end{equation}
which is correct for anomaly cancellation in type I on the quintic.

To complete the story, we note that the naive result \eqref{naive},
$4(-7+5H^2)$ differs from the type I result with $B=0$ simply by one unit
of D6-brane charge, as well as a sign. Both can be understood by
noting that the path starting at large volume with $B=0$ has to go
through the conifold singularity before reaching the LG point. At
the conifold, the O-plane looses exactly one unit of the vanishing
cycle, which is the D6-brane, and also changes into an anti-orientifold
plane, see \cite{g2}. We thus see that we can understand completely the
charge of the orientifold plane under analytical continuation through
the quintic moduli space, and that large volume and Gepner/LG data agree
beautifully.

\subsubsection{Exchange orientifolds}

Consider now the orientifold action
\begin{equation}
\eqlabel{sig1}
\sigma_1 :(x_1,x_2,x_3,x_4,x_5)\to (-x_2,-x_1,-x_3,-x_4,-x_5)\,.
\end{equation}
Its LG charges are
\begin{equation}
\langle C|k\rangle = (1+\om^{3k})(1-\om^{3k})^4\,,
\end{equation}
which gives at large volume
\begin{equation}
\begin{split}
[O_1] &=  -1+2H-\frac{19}{15}H^3\,, \\
g^2[O_1] &= -2 H +H^2 + \frac{16}{15}H^3\,.
\end{split}
\end{equation}
Again, we can check that this matches the geometric expectations.
At large volume, the fixed point locus of the involution consists
of two components \cite{BH2}: An O7-plane at a hyperplane $x_1=x_2$,
and an O3-plane at a point $x_1=-x_2$, $x_3=x_4=x_5=0$. The general
formulas (see, \eg, \cite{bhhw}) give the O-plane charge of a fixed
component $Y\subset X$ as
\begin{equation}
\pm[Y]\; \frac{2^{3-{\rm codim}_{\reals}(Y)}}{\sqrt{\widehat A(X)}}\;
\sqrt{\frac{L\bigl(\frac 14 T Y\bigr)}{L\bigl(\frac 14 N Y\bigr)}}\,,
\end{equation}
where $[Y]$ is the Poincar\'e dual of the fixed point locus, and the
sign $\pm$ is the type of O-plane (O$^{+}$ or O$^{-}$). For the quintic
$X$ in $\projective^4$, $\widehat A(X)= 1+\frac{10}{12}H^2$. The hyperplane
has $[Y]=H$, and $c(NY)=1+H$, so $c(TY)=(1+10H^2-40H^3)/(1+H)=1-H+11H^2$.
We find $L\bigl(\frac14 NY\bigr)= 1+H^2/48$, $L\bigl(\frac 14 TY\bigr)=
1-21 H^2/48$. For the point on the quintic, $[pt]=H^3/5$, so the formula
evaluates altogether to
\begin{equation}
\pm 2 H \sqrt{\frac{(1-21H^2/48)}{(1+H^2/48)(1+10 H^2/12)}}
\pm \frac{H^3}{40} = \pm \bigl(2H- \frac{31}{24} H^3\bigr) \pm
\frac{H^3}{40}\,.
\end{equation}
Let's compare this with $[O_1]$ and $g^2[O_1]$ we have computed above.
First of all, we have to add $1$ to $[O_1]$ because the path to large
volume crosses the conifold locus. Then we see that the resulting
O-plane is an O7 with an O3 of the same type (we can't determine the
overall type from these considerations)
\begin{equation}
2H-\frac{31}{24} H^3 + \frac{H^3}{40} = 2H -\frac{19}{15}H^3\,.
\end{equation}
For $g^2[O_1]$, we have to multiply it with $\ee^{-H/2}$ because of the
B-field, and find that this is an O7 with an O3 of the opposite type
\begin{equation}
\ee^{-H/2} \bigl(2H-\frac{31}{24} H^3 - \frac{H^3}{40}\bigr)
= 2H + H^2- \frac{16}{15} H^3\,.
\end{equation}
It should be possible to understand geometrically why the B-field changes
the type of the O3-plane in this fashion.

Finally, we consider the orientifold action with two exchanges, which
is in LG limit:
\begin{equation}
\sigma_2 :(x_1,x_2,x_3,x_4,x_5)\to (-x_2,-x_1,-x_4,-x_3,-x_5)\,.
\end{equation}
Its LG charges are
\begin{equation}
\langle C|k\rangle = (1+\om^{3k})^2(1-\om^{3k})^3
\end{equation}
which gives at large volume
\begin{equation}
\begin{split}
[O_2] &= 1 - \frac{3}{5} H^2\,, \\
g^2[O_2] &= \frac{2}{5}H^2 - \frac{1}{5}H^3\,.
\end{split}
\end{equation}
In the geometric regime, the fixed point locus corresponds to an O5 at
a degree 5 curve at $x_1=x_2$, $x_3=x_4$, $2x_1^5+2x_3^5+x_5^5=0$ in
cohomology class $H^2$, plus an O5 at a rational curve $x_1=-x_2$,
$x_3=-x_4$, $x_5=0$ in class $H^2/5$. The general formula evaluates to
\begin{equation}
\pm \frac12 \frac{H^2}{5} \pm \frac 12 H^2\,.
\end{equation}
Indeed, removing the $1$ from $[O_2]$, this is
\begin{equation}
-\frac35 H^2 = -\frac 12H^2 -\frac 1{10}H^2\,,
\end{equation}
whereas for $g^2[O_2]$, we get
\begin{equation}
\frac 25H^2-\frac15 H^3= \ee^{-H/2} ( \frac 12H^2-\frac 1{10}H^2)\,.
\end{equation}
Again, the type of one component of the O-plane changes as we navigate
through the non-geometric phase, or as we change the B-field from
$0$ to $1/2$.

\end{document}